%% file: main.tex
\documentclass[twocolumn,amsfonts, amssymb, amsmath, nobibnotes, aps, prd, nofootinbib, superscriptaddress]{revtex4-2}

\usepackage{amsmath}
\usepackage{amsfonts}
\usepackage{amssymb}
\usepackage{booktabs}
\usepackage[T1]{fontenc}
\usepackage{graphicx}
\usepackage[colorlinks=true, linkcolor=blue, urlcolor=blue, citecolor=blue]{hyperref}
\usepackage{siunitx}
\usepackage{xcolor}
\usepackage{orcidlink}

\usepackage{graphicx}	% Including figure files
\usepackage{amsmath}	% Advanced maths commands
\usepackage{amssymb}	% Extra maths symbols
\usepackage{siunitx}
\usepackage{xcolor}

\graphicspath{{figures/}}

\newcommand{\uib}{IAC3, Universitat de les Illes Balears, Cra. de Valldemossa km 7.5, 07122 Palma, Spain}
\newcommand{\gaia}{\textit{Gaia}}
\newcommand{\fastt}{\texttt{FastTracks}}
\newcommand{\jax}{\texttt{JAX}}

\newcommand{\depth}{\mathcal{D}}
\newcommand{\F}{\mathcal{F}}

\newcommand{\pyfstat}{\texttt{pyfstat}}
\newcommand{\ptemcee}{\texttt{ptemcee}}
\newcommand{\Sn}{S_{\mathrm{n}}}

\newcommand{\Tobs}{T_{\rm obs}}
\newcommand{\Tsft}{T_{\rm SFT}}

\begin{document}

\title{Search Method for Continuous Gravitational Waves from Rotating Neutron Stars in Highly Eccentric Wide Orbits}

\author{M.~Ball\orcidlink{0000-0001-5565-8027}}
\email{matthew.ball@ligo.org}
\affiliation{\uib}

\author{S.~Bussières\orcidlink{0009-0002-8990-4788}}
\email{sofia.bussieres@ligo.org}
\affiliation{\uib}

\author{L.~Mirasola\orcidlink{0009-0004-0174-1377}}
\email{lorenzo.mirasola@ligo.org}
\affiliation{\uib}

\author{J.~R.~Merou\orcidlink{0000-0002-5776-6643}}
\affiliation{\uib}

\author{A.~K.~Sharma\orcidlink{0000-0003-0067-346X}}
\affiliation{\uib}

\author{N.~Singh\orcidlink{0000-0002-1135-3456}}
\affiliation{\uib}

\author{D.~Keitel\orcidlink{0000-0002-2824-626X}}
\affiliation{\uib}

\author{A.~M.~Sintes\orcidlink{0000-0001-9050-7515}}
\affiliation{\uib}

\begin{abstract}
    Recent results from the \gaia{} third data release have highlighted 21 potential neutron stars candidates in wide, highly-eccentric binary systems and provide estimates of their orbital parameters.
    We present a feasibility study for a directed search for continuous gravitational waves (CWs) tailored for these and similar systems.
    To this extent, we have computed a frequency metric to place templates in the parameter space accounting for uncertainties on the orbital parameters.
    We have built upon the \fastt{} infrastructure to efficiently tackle the high-dimensional search space, leveraging a combination of downsampling and interpolation to reduce the computational cost of generating the binary Doppler correction.
    We performed an injection campaign to demonstrate the performance of this procedure and showed that a search using this approach is viable.
    Such a search would represent the first combination of \gaia{} observations and CW analysis. 
\end{abstract}

\maketitle

\section{Introduction}
\input{introduction}

\section{Signal Model}
\label{sec:waveform}
\input{signal_model}

\section{Parameter space and data}
\label{sec:param_space}
\input{data.tex}

\section{Dealing with highly eccentric orbits}
\label{sec:metric}
\input{template_spacing.tex}

\section{Search Algorithm}
\label{sec:search_method}
\input{search_algorithm.tex}

\section{Specific Implementation}
\label{sec:implementation}
\input{implementation.tex}

\section{Injections}
\label{sec:injections}
\input{injections.tex}

\section{Conclusion}
\label{sec:conclusion}
\input{conclusion.tex}

\section*{Acknowledgements}

We thank Francesco Amicucci and the CW working group of the LIGO--Virgo--KAGRA Collaboration for comments on the manuscript.

This material is based upon work supported by NSF's LIGO Laboratory which is a major facility fully funded by the National Science Foundation. The authors are grateful for computational resources provided by the LIGO Laboratory and supported by National Science Foundation Grants PHY-0757058 and PHY-0823459.

This work was supported by the Universitat de les Illes Balears (UIB) with funds from the Programa de Foment de la Recerca i la Innovació de la UIB 2024-2026 (supported by the yearly plan of the Tourist Stay Tax ITS2023-086); the Spanish Agencia Estatal de Investigación grants PID2022-138626NB-I00, PID2025-170644NA-I00, RED2024-153978-E, RED2024-153735-E, funded by MICIU/AEI/10.13039/501100011033 and the ERDF/EU; and the Comunitat Autònoma de les Illes Balears through ``Ajuts per a projectes de recerca científica i tecnològica'' (PRD2025\_099 and PRD2025\_100) and through the Conselleria d'Educació i Universitats with funds from the European Union - European Regional Development Fund (ERDF) (SINCO2022/18146 - Plataforma HiTech-IAC3-BIO); and by COST action SCALES CA24139, supported by COST (European Cooperation in Science and Technology).

SB is supported through the Conselleria d'Educació i Universitats del Govern de les Illes Balears via an FPI-CAIB doctoral grant (FPI2025\_030) with funds from the European Social Fund+ in the framework of the Balearic Islands ESF+ Program 2021-2027.

JRM is supported by the Spanish Ministerio de Ciencia, Innovación y Universidades (ref. FPU22/01187).

NS is supported through the Conselleria d'Educació i Universitats del Govern de les Illes Balears via an Vicenç Mut postdoctoral grant (POSTDOC2024\_55) with funds from the European Social Fund+ in the framework of the Balearic Islands ESF+ Program 2021-2027.

The authors gratefully acknowledge the supercomputing resources provided by the Balearic Islands Centre for Supercomputing and AI (BSAI). Significant portions of the analysis were run in the GPP and ACC partitions of the TalaIA supercomputer.

The authors gratefully acknowledge the computer resources at Artemisa, funded by the European Union ERDF and Comunitat Valenciana as well as the technical support provided by the Instituto de Fisica Corpuscular, IFIC (CSIC-UV).

This paper has been given LIGO DCC number LIGO-P2600594.

\appendix
\section{Derivatives of the binary Doppler shift}
\label{app:orbital_derivatives}
\input{appendix_derivatives}

\section{Additional injections at high frequencies}
\label{app:high_ecc_inj}
\input{appendix_addtl_injections.tex}

\bibliography{references}

\end{document}

%% file: introduction.tex
Continuous gravitational waves (CWs) are faint, persistent signals canonically
associated to non-axisymmetric rotating neutron stars (NSs)~\cite{Riles:2022wwz}.
Although still undetected, a discovery would provide information about the internal
structure of the star and on the NSs distribution in our galaxy~\cite{2021ApJ...921...89R, 2005AJ....12ss9.1993M, Prabhu:2024lzl}.
These asymmetries in the mass distribution (``mountains'') are theorized to be due to elastic stresses on the crust~\cite{2024CQGra..41d3001G} 
or internal deformations due to strong magnetic fields~\cite{1996A&A...312..675B,2002PhRvD..66h4025C}.
The primary emission models include single harmonic emission at exactly twice the rotation rate~\cite{Riles:2022wwz,2023APh...15302880W,2023NatAs...7.1160H} 
and dual harmonic emission at one and two times the rotation rate~\cite{2010MNRAS.402.2503J}. 
Additional emission mechanisms include r-modes induced by internal fluid oscillations~\cite{1998ApJ...502..708A} 
and quasi-periodic emission induced by free precession due to a misalignment between the rotation axis and the axis of symmetry~\cite{2005CQGra..22.1825V}.

CWs are expected to be much weaker signals compared to those currently observed by
the LIGO-Virgo-KAGRA (LVK) network~\cite{KAGRA:2013rdx,LIGOScientific:2014pky,VIRGO:2014yos,10.1093/ptep/ptaa125}.
As a consequence, CW searches integrate over year-long timespans to accumulate
signal-to-noise-ratio to achieve the detection.
Due to this, they have to account for long-term evolution of the emitted signal (see, e.g.,~\cite{Tenorio:2021wmz, Piccinni:2022vsd}).
More specifically, they have to consider the change of the spin frequency over time
(the so called spin-down effect) and the Doppler modulation induced by the Earth's motion 
and by orbital modulation for sources in binary systems.
Searches for a CW emitted by the latter can be potentially challenging, as the orbital motion
can depend on up to five additional parameters.
Therefore, the complexity of these searches can be significantly higher compared to those for isolated
NSs, requiring a tailored approach to cover the orbital parameter space~\cite{Leaci:2015bka, Mukherjee:2022tuc, Goetz:2011bd, Covas:2019jqa, Covas:2022mqo, Tenorio:2024jgc, Singhal:2019dfn, Amicucci:2025aiz} unless orbital parameters are well constrained (see, e.g.,~\cite{PhysRevD.110.123043, Mirasola:2025car, LIGOScientific:2026qsb}).

All-sky searches for CWs from NSs in binaries explore unknown orbital parameters in addition to sky position and 
frequency parameters, making them significantly more expensive than other CW searches.
Due to computational constraints, usual searches focus on circular orbits, remaining sensitive to only small eccentricities
(see, e.g.,~\cite{LIGOScientific:2020qhb, Covas:2019jqa, Covas:2024nzs, LIGOScientific:2025ouy, Covas:2026raf}).

The \gaia{} third data release of \gaia{} (DR3)~\cite{2023A&A...674A..34G}, however, provides observational motivation for extending search methods to higher values.
Through spectroscopic and radial-velocity follow-up of \gaia{} DR3 astrometric sources, Ref.~\cite{2024OJAp....7E..58E} identified $21$ Gaia DR3 astrometric binaries containing solar-type main-sequence stars and unseen companions consistent with NSs. These systems have orbital periods of approximately $190-1050\,\mathrm{days}$ and eccentricities reaching $e\simeq0.8$. 
Most lie at sub-kiloparsec distances, with the farthest candidate at approximately $1.5\,\mathrm{kpc}$. 
Although the spins of these potential NSs remain unknown, the excellent sky localization accuracy and known orbital parameters of these systems can constrain the search volume, making these high-eccentricity candidates viable targets for directed CW searches.
If accomplished, a search of this kind would represent a first application of \gaia{} observations for a CW search.

Given the unexplored parameter-space region of these systems, we developed a ``frequency metric'' following a similar approach to that detailed in Ref.~\cite{Tenorio:2024jgc}, allowing us to place templates within the uncertainties provided by \gaia{} observations.
However, this initial step does not account for correlations among parameters, which could be incorporated through a yet-to-be-derived parameter-space metric.
Such a metric would complement the one derived in Ref.~\cite{Leaci:2015bka} for low-eccentricity systems.

In this work, we explore the feasibility of a search tuned for these systems using \fastt{}~\cite{Tenorio:2024jgc},
which is a pipeline built on the highly parallelized \jax{} engine~\cite{Frostig2018,jax2018github}.
It is designed to evaluate detection statistics for generic CW signals on Short Fourier Transform (SFT)-like data~\cite{Allen:T040164}.

For highly-eccentric systems, the binary Doppler computation represents a limiting factor due to its intrinsic transcendental nature.
We show how, for wide-orbit systems, this step can be significantly sped up through downsampling and interpolating to the required resolution, balancing computing cost and accuracy.

As an additional novelty of this work, we place templates using a hybrid approach to optimize the usage of \gaia{} observations in the CW search stage, combining a \fastt{}-like random template placement with a grid-based approach.

The paper is organized as follows. 
We discuss the assumed signal model in Section~\ref{sec:waveform}.
Section~\ref{sec:param_space} describes the \gaia{} and gravitational wave (GW) data involved in this study.
We introduce the frequency metric for the high-eccentricity regime in Section~\ref{sec:metric}, and we give an estimation of computing cost for a CW analysis dedicated to the \gaia{} DR3 NS candidates from~\cite{2023A&A...674A..34G}.
Section~\ref{sec:search_method} shows the global approach in the \fastt{} infrastructure, and the implementation of a tailored methodology for these targets is detailed in Section~\ref{sec:implementation}.
We demonstrate the pipeline in action through injection recovery using simulated and real data from the second part of the LVK's fourth observing run (O4b) in Section~\ref{sec:injections}.
Finally, we summarize our results in Section~\ref{sec:conclusion}.

%% file: signal_model.tex
For our analysis, we will consider the case of single-harmonic CW emission at two times the NS rotation frequency.
The CW signal at the detector can be expressed as
\begin{align}
    \label{Eq:hSignal}
    h(t)  = h_0\,F_+(t;\hat{n},\psi)\,  \frac{1+\cos^2\iota}{2}\, \cos \phi(t) +\nonumber\\
    h_0\,F_\times(t;\hat{n},\psi)\, \cos\iota\,\sin \phi(t), 
\end{align}
where, $\phi$ represents the GW phase, modulated by the relative motion of the source and detector; $\iota$ is the inclination angle of the NS rotational axis with respect to the line of sight; $F_{+, \times}$ are the time-dependent beam pattern functions~\cite{Jaranowski:1998qm} that encode the detector response as a function of the source sky location $\hat{n}$ and of the polarization angle $\psi$.

The strain amplitude $h_0$ is given by~\cite{Riles:2022wwz}
\begin{equation}
    h_0 = \frac{4\pi^2G}{c^4}\frac{I_{zz}\epsilon}{d}f_0^2\,,
    \label{eq:strain_amplitude}
\end{equation}
where $f_0$ is the GW frequency and $\epsilon$ is the equatorial ellipticity, given by
\begin{equation}
    \epsilon \equiv \frac{|I_{xx} - I_{yy}|}{I_{zz}}\,,
    \label{eq:ellipticity_inertia}
\end{equation}
with $I_{zz}$, $I_{xx}$, and $I_{yy}$ the principal moments of inertia for a star assumed to be rotating about the z-axis.
The ellipticity can also be related to the mass quadrupole moment $Q_{22}$ of the source via \cite{10.1046/j.1365-8711.2000.03938.x}
\begin{equation}
    \epsilon = \sqrt{\frac{8\pi}{15}}\frac{|Q_{22}|}{I_{zz}}.
    \label{eq:ellipticity_quadrupole}
\end{equation}

The GW frequency is modeled as nearly monochromatic, slowly decreasing due to the spin-down of the NS.
This signal is modulated by a Doppler shift due to the relative motion of the source and the detector.
Up to the first-order contribution, the combined frequency evolution can be written as~\cite{Edwards:2006zg,Leaci:2016oja}
\begin{equation}
    f(t) = \frac{d\phi(t)}{dt} \approx \left(f_0 + f_1\Delta t\right)\left[1+\frac{\vec{v}}{c}\cdot\hat{n}-\frac{\dot{R}}{c}\right]
    \label{eq:ecc_freq_evolution}
\end{equation}
where $f_0$ is the frequency at the reference time $t_{\rm{ref}}$, $\Delta t$ is the time elapsed from the reference time, $f_1$ is the spin-down rate, $\vec{v}$ is the detector's velocity vector relative to the  Solar System Barycenter (SSB), and $\hat{n}$ is the unit vector pointing from the SSB to the Binary Barycenter (BB).
$\dot{R}/c$ describes the effect of the binary motion around the BB, given by~\cite{Covas:2019jqa}

\begin{equation}
    \frac{\dot{R}}{c} = \frac{a_p \left(\cos\omega\cos E\sqrt{1-e^2} - \sin\omega\sin E\right)}{\frac{P_{\mathrm{orb}}}{2\pi}\left(1-e\cos E\right)}\,, 
    \label{eq:Rdot_over_c}
\end{equation}
where $a_p$ is the projected semi-major axis of the binary orbit, $P_{\mathrm{orb}}$ is the orbital period, $e$ is the orbital eccentricity, 
$\omega$ is the argument of periastron, and $E$ is the time-dependent eccentric anomaly, computed numerically from the transcendental equation 
\begin{equation}
    t-t_p = \frac{P_{\mathrm{orb}}}{2\pi}\left(E-e\sin E\right)\,,
    \label{eq:ecc_anomaly}
\end{equation}
where, $t_p$ is the time of the periapse's passage.

%% file: data.tex
We here describe the datasets used in this work. Section~\ref{sec:search_parameter_space} characterises the NS candidates identified in the \gaia{} DR3 by Ref.~\cite{2024OJAp....7E..58E}, highlighting how the binary systems in which they lie are outside parameter spaces covered by present CW searches. Successively, Section~\ref{sec:data} focuses on the GW detectors' data used in the context of this work to assess the feasibility of a CW search dedicated to these new targets.

\subsection{\gaia{} candidates parameter space\label{sec:search_parameter_space}}

\begin{figure*}[!ht]
    \centering{}
    \includegraphics[width=\textwidth]{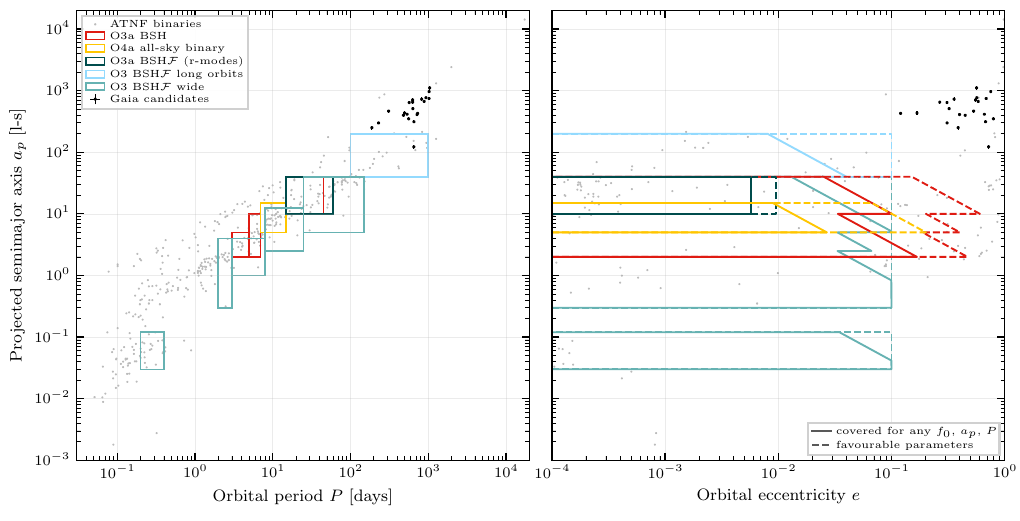}
    \caption{\textit{Left panel:} $P_{\rm orb}$ and $a_p$ of the 21 \gaia{} NS candidates, compared with the regions covered by previous all-sky searches for CWs from NSs in binary systems (colored boxes): the O3a \textsc{BinarySkyHough} (BSH) search~\cite{LIGOScientific:2020qhb}, the O4a all-sky binary search~\cite{LIGOScientific:2025ouy}, and the \textsc{BinarySkyHou}$\F$ (BSH$\F$) searches~\cite{Covas:2022rfg}, for long orbital periods~\cite{Covas:2024nzs}, and over a wide parameter space~\cite{Covas:2026raf}. Black dots are the 21 \gaia{} targets and grey dots the ATNF binaries~\cite{2005AJ....12ss9.1993M}. \textit{Right panel:} $e$ and $a_p$ covered by the discussed searches. Solid contours mark the eccentricity covered by the respective coloured boxes in the left panel, and dashed lines the eccentricity range for the most favourable parameters, that is, at the lowest $f_0$, the smallest $a_p$ and the longest $P_{\mathrm{orb}}$. \label{fig:orbital_parameter_space}}
\end{figure*}

The \gaia{} DR3, based on observations between July 2014 and May 2017, and released on June 13 2022, contains approximately 1.81~billion sources, predominantly stars. The binary-star catalog includes 169,227 candidates with known astronometric orbital solutions~\cite{2023A&A...674A..34G}, allowing for searches even for dark companions in binaries. Ref.~\cite{2023MNRAS.518.2991S} identified 177 binaries with probable compact companions. Building on this catalogue and additional \gaia{}-selected candidates, Ref.~\cite{2024OJAp....7E..58E} combined multi-epoch radial-velocity measurements with \gaia{} astrometry to identify 21 binaries whose unseen companions are consistent with NSs. In this study, we focus on these 21 \gaia{} DR3 binaries. 

The joint astrometric and spectroscopic fits constrain the orbital period, eccentricity, inclination, time of periastron passage, and argument of periapsis \cite{2023A&A...674A..34G,2024OJAp....7E..58E}. The projected semi-major axis of the NS is derived from that of the companion and from the mass ratio as 
\begin{equation} 
    a_p = a_{p,\mathrm{c}}\frac{M_{\mathrm{c}}}{M_{\mathrm{NS}}},{}
\end{equation}
with companion masses between $0.78$ and $1.32\,M_\odot$ and compact-object masses between $1.26$ and $1.90\,M_\odot$. The resulting orbits have periods between 189 and 1046~days and projected NS semi-major axes between 123 and 1106 light-seconds. The details of all candidate NSs' orbital parameters obtained through the radial velocity fitting and \gaia{} astrometry can be found in Table~4 of Ref.\cite{2024OJAp....7E..58E}.

These values place the targets outside of the parameter regions covered by previous all-sky searches for CWs from NSs in binary systems, as shown in the left panel in Figure~\ref{fig:orbital_parameter_space}. The parameter space coverage for binary all-sky searches is generally limited by the computational cost associated with the unknown sky positions, orbital parameters, frequencies, and spin-down terms. For the \gaia{} targets, the constrained orbital parameter space and the well-known sky positions substantially reduce the required parameter space coverage. However, the orbital parameter uncertainties still need to be accounted for in the proposed search. 

Moreover, the discussed all-sky searches were designed to detect signals from NSs in circular orbits. The circular orbit template banks used in these previous searches can also recover mildly eccentric signals, remaining sensitive only up to a maximum eccentricity, set by requiring that the eccentric track stay within one frequency bin of the circular track within the observation period~\cite{LIGOScientific:2020qhb, Covas:2024nzs, LIGOScientific:2025ouy}. In contrast, the \gaia{} NS targets have orbits that are significantly eccentric, with $e$ between $0.12$ and $0.80$. The right panel in Figure~\ref{fig:orbital_parameter_space} compares the eccentricity range of these previous searches with the eccentricities of our targets. Even the single target in the left panel of \mbox{Figure~\ref{fig:orbital_parameter_space}} whose $(P_{\mathrm{orb}}, a_p)$ falls inside their previously searched values in Ref.~\cite{Covas:2024nzs} has $e=0.72$, much greater than that search's maximum covered eccentricity of $0.1$. A circular waveform model is therefore not useful for our targets. Hence, we use eccentric-orbit waveform model detailed in Section~\ref{sec:waveform} and design our search methodology to remain sensitive over the ranges of eccentricity and argument of periapsis of the \gaia{} NS targets explicitly.

For these NS candidates, the simplest evolutionary scenarios suggest that the compact objects have not undergone accretion-driven spin-up \cite{2024OJAp....7E..58E}. In the absence of significant recycling, their present-day rotational frequencies are expected to be determined primarily by their birth spin periods and subsequent spin-down, which may result in rotational frequencies below those typically observed in recycled millisecond pulsars. Although the spin frequencies of these sources remain unknown, this evolutionary picture supports focusing the proposed search on relatively low CW frequencies. We then focus on our studies in the frequency range of $50-200$~Hz. This choice represents a compromise between astrophysical motivation, detector sensitivity, and computational feasibility, rather than a prediction of the candidates' actual spin frequencies. The detector sensitivity and computational considerations that informed the choice of frequency range are discussed in Sections~\ref{sec:data} and \ref{sec:template_density}, respectively.

\subsection{Data\label{sec:data}}

We use GW data from the O4b run of the LVK network~\cite{LIGOScientific:2026jgl}. The run spans from 10 April 2024 15:00 UTC (GPS time 1396796418 s) to 28 January 2025 17:00 UTC (GPS time 1422118818 s). 

We use data from the two LIGO detectors at Hanford (H1) and Livingston (L1)~\cite{LIGOScientific:2014pky}. Virgo~\cite{VIRGO:2014yos} was observing during O4b, but its inclusion would not appreciably improve the sensitivity of a search in the considered frequency band~\cite{VIRGO:2026okv} while increasing its computational cost. KAGRA~\cite{10.1093/ptep/ptaa125} data are likewise not used. During O4b, H1 and L1 were in observing mode 48.6\% and 68.1\% of the time, respectively~\cite{LIGOScientific:2026jgl, PhysRevD.111.062002, LIGO:2026ika}.

We produce our own SFTs~\cite{Allen:T040164} using LALSuite~\cite{lalsuite}. We start from the C00 calibrated strain channel, self-gated to remove large transient disturbances in the time domain~\cite{Davis:T2400003, LIGO:2026ika}, \texttt{GDS-CALIB\_STRAIN\_CLEAN\_AR}, and restrict it to the analysis-ready science segments with CAT1 vetoes applied~\cite{Goetz:T2400058}, trimmed to the O4b times. 

The SFTs are computed with a duration of $\Tsft = 5000$~s. This duration is consistent with the signal power of our targets remaining within a single frequency bin over each SFT. With these data-quality criteria and $5000$~s SFT duration requirement, the SFTs span a shorter interval than the full O4b run, from 12 April 2024 15:00 UTC (GPS time 1396969218~s) to 28 January 2025 15:45 UTC (GPS time 1422114318~s). We are left with 2020 SFTs for H1 and 3013 SFTs for L1, equivalent to 116.9 and 174.4 days of data, respectively.

The data contain narrow spectral artifacts, such as instrumental lines and combs, which are persistent, narrow-band features of non-astrophysical origin that can mimic the behavior of a CW signal~\cite{Jaume:2024eob, O4LIGODetector:2026okh,2018PhRvD..97h2002C}. These artifacts can be identified and mitigated by using a list of known lines~\cite{Goetz:T2500212} and combs, to apply vetoes to candidates that are affected by them. These narrow-band artifacts are also predominantly abundant at lower frequencies, in addition to lower detector sensitivity in the low frequency range~\cite{LIGOScientific:2026jgl}. We therefore set the lower frequency cutoff of this study at~$50$ Hz. The upper cutoff of $200$~Hz is motivated by computational constraints and the unrecycled nature of the \gaia{} NS candidates, as discussed in Sec.~\ref{sec:search_parameter_space}. We generate SFTs over the slightly wider frequency range of $49.5$--$200.5$~Hz.

%% file: template_spacing.tex
\gaia{} observations constrain the orbital parameters of the candidate systems, defining the parameter-space region that could be searched. However, their high eccentricities limit the applicability of the standard low-eccentricity metric~\cite{Leaci:2015bka}.

In Section~\ref{sec:resolution} we describe how to determine the parameter-space resolution suitable for these systems. We then use it to estimate the number of templates required to cover each target's orbital-parameter uncertainties in Section~\ref{sec:template_density}. 

\begin{figure*}[!ht]
    \centering
    \includegraphics[width=\textwidth]{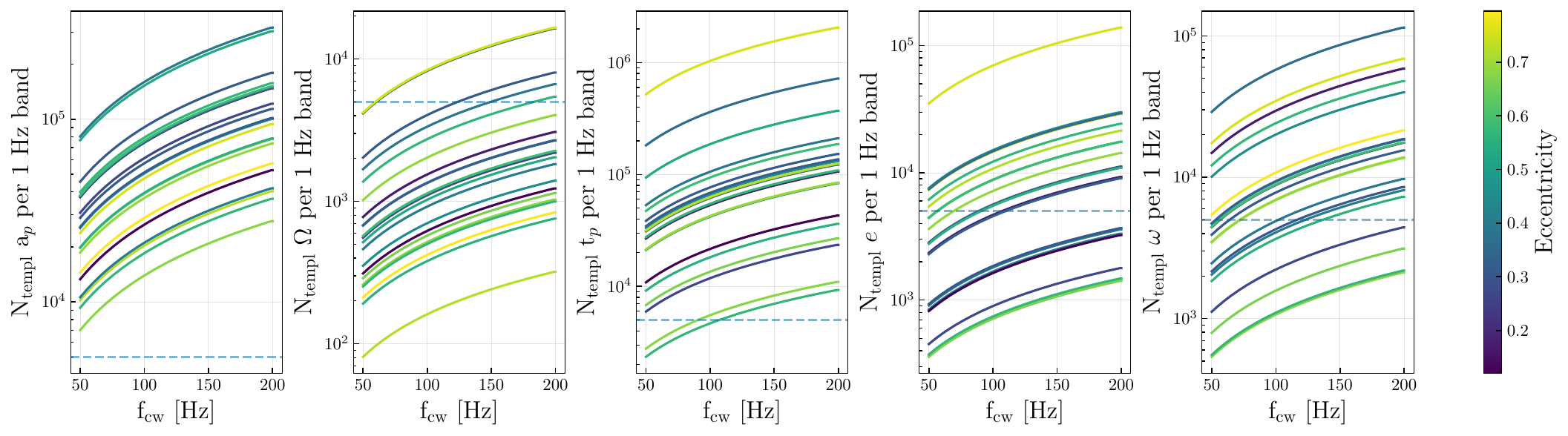}
    \caption{Per-parameter number of templates as a smooth function of frequency integrated over a 1~Hz band for each of the \gaia{} targets. Plots are colored by the binary system's eccentricity. The largest contributions tend to be from the time of periastron's passage $t_p$, and the smallest contribution is from the orbital frequency $\Omega$. Note that this only accounts for $\pm3\sigma$ uncertainties as provided in the DR3 catalog. The blue dotted lines correspond to $N_\mathrm{templ} = 1$ for all individual frequency bins in the 1~Hz band, so a curve below the line indicates that the uncertainty in that parameter is narrower than the template spacing at the given frequency.}
    \label{fig:num_templates}
\end{figure*}

\subsection{Parameter-space resolution\label{sec:resolution}}

To determine the template spacing, we quantify how offsets in the search parameters change the predicted instantaneous frequency of a signal. Our approach uses SFTs, each computed over a segment of duration $\Tsft$, giving a frequency-bin width of $\delta f=1/\Tsft$.
Following the same criterion as \fastt{}~\cite{Tenorio:2024jgc}, we require that the frequency tracks of two neighboring templates, separated by $\delta\lambda$ in a single parameter $\lambda$, differ a frequency bin at any time during the observation,
\begin{equation}
    \max_{t}\left|\frac{\partial f(t;\lambda)}{\partial\lambda}\right|\delta\lambda = \delta f\,,
    \label{eq:resolution_condition}
\end{equation}
which can be inverted to give the resolution spacing.

The signal model depends on the parameters $\lambda\in \{f_0, f_1, \alpha, \delta, a_p, \Omega, t_p, e, \omega\}$, where $\Omega \equiv 2\pi/P_{\mathrm{orb}}$ is the mean orbital angular frequency. The precision of \gaia{} astrometry allows us to fix the sky position of the targets, given by their right ascension and declination  $\{\alpha, \delta\}$, to their catalog values propagated to the observing epoch. We compute the resolution equations separately for each remaining parameter. This effort was already undertaken in Ref.~\cite{PhysRevD.110.123043}. Here we rederive these equations and correct the equations presented in Appendix~A of that reference. 

Applying the inverted Equation~\ref{eq:resolution_condition} to the frequency model in Equation~\ref{eq:ecc_freq_evolution}, we adopt the following grid spacings for the intrinsic GW frequency and spindown
\begin{align}
    \delta f_0 &= \frac{1}{\Tsft}\,, \label{eq:delta_f0}\\
    \delta f_1 &= \frac{1}{\Tsft\,\Tobs}\,. \label{eq:delta_f1}
\end{align}
where we have neglected the small Doppler corrections to $\partial f/\partial f_0$ and $\partial f/\partial f_1$ and take $\max|\Delta t|=\Tobs$, the whole observation time.

The orbital parameters $\{a_p,\Omega,t_p,e,\omega\}$ affect the observed frequency through the binary Doppler shift $\dot R/c$. 
The frequency evolution of the \gaia{} NS candidates is expected to be dominated by their intrinsic spindown.
Inspired by all-sky searches for isolated NSs (e.g.,~\cite{LIGOScientific:2025ouy}), we focus our preliminary studies in range $f_1\in[-10^{-9},0]$~Hz/s, the change in instantaneous frequency over the observing time  is $|f_1|\,\Tobs \ll f_0$. We therefore approximate
\begin{equation}
    \frac{\partial f}{\partial\lambda} \simeq -f_0\,\frac{\partial(\dot R/c)}{\partial\lambda}\,.
    \label{eq:df_dlambda_approx}
\end{equation}

To evaluate these derivatives, we account for both the explicit parameter dependence of $\dot R/c$ in Equation~\ref{eq:Rdot_over_c} and its implicit dependence through the eccentric anomaly $E(t)$, determined by Kepler's
equation in Equation~\ref{eq:ecc_anomaly}. The resulting expressions can be found in Appendix~\ref{app:orbital_derivatives}.
Inserting these derivatives into Equation~\ref{eq:resolution_condition} gives
\begin{equation}
    \delta\lambda = \left[f_0\Tsft\,\max_{T_{\rm obs}}\left|\frac{\partial(\dot R/c)}{\partial\lambda}\right|\right]^{-1},\quad \lambda\in\{a_p,\Omega,t_p,e,\omega\}\,,
    \label{eq:delta_orbital}
\end{equation}
which can be evaluated for each of the \gaia{} targets over the considered observation period $T_{\rm obs}$.

\subsection{Expected number of templates\label{sec:template_density}}

For each NS target, we compute the resolution for each binary parameter and compute the number of templates that fit within the uncertainty range estimated by \gaia{}. Since the true parameters may lie outside the $1\sigma$ region, we choose our template bank to cover the $\pm3\sigma$ window for each parameter. We compute the total number of templates $\rm{N}_{\rm{templ}}(\lambda) = 6\sigma/\delta\lambda$ needed to cover the parameter space for each binary parameter via Equation~\ref{eq:delta_orbital} and report the number of templates within the frequency range described in Section~\ref{sec:search_parameter_space} in Figure~\ref{fig:num_templates}. The number of templates per frequency bin contributed by each binary parameter ranges from less than 1 template for some binaries to close to 400 for other targets. Some binary parameters contribute far more templates than others; for instance, the orbital period term $\Omega$ rarely contributes more than a handful of templates, but the time of periastron passage term $t_p$ can contribute hundreds. For most candidates, the number of templates is dominated by the $t_p$ contribution, which strongly impacts the phase of the binary orbit. 

\begin{figure}[!ht]
    \centering
    \includegraphics[width=\columnwidth]{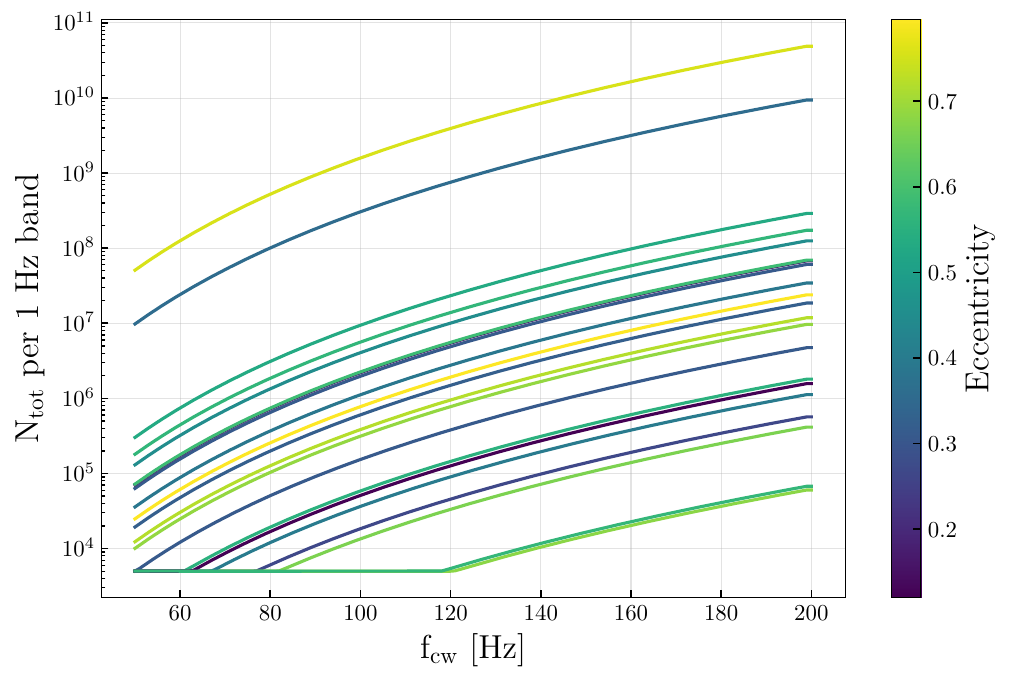}
    \caption{Total number of binary templates required per 1~Hz band for each \gaia{} target we consider. Plots are colored by the binary system's eccentricity. Note that this does not include templates due to the frequency derivative $f_1$.}
    \label{fig:tot_num_templates}
\end{figure}

Multiplying the per-parameter contributions gives the total number of templates, shown in Figure~\ref{fig:tot_num_templates}. In this calculation, cases where the per-frequency bin ($1/Tsft$) $N_{\rm{templ}}<1$ are set to 1.
This estimation, does not account for templates in the $f_1$ space which would be multiplicative with those curves.
Considering $f_1\in[-10^{-9},0]$~Hz/s, we obtain for the majority of targets, a total number of templates comparable with those in Table 2 of Ref.~\cite{LIGOScientific:2026ppd}.

%% file: search_algorithm.tex
As highlighted in Section~\ref{sec:metric}, some of these interesting targets require a significant computational investment due to the large number of templates needed to cover the parameter space. In addition, the heterogeneity of the targeted binary systems (see, e.g., the eccentricities spanned in Figure~\ref{fig:orbital_parameter_space}) requires a tailored approach for each of them. Therefore, we need a search algorithm that is both flexible and efficient. In this work, we use \fastt{}~\cite{Tenorio:2024jgc} as an algorithm that meets these requirements. 

\fastt{} evaluates the detection statistic for individual templates and is accelerated through the use of Graphics Processing Units (GPUs). After an initial search stage, interesting candidates are selected for follow-up analyses. This Section is divided as follows: Section~\ref{subsec:fasttracks} gives an overview of the \fastt{} algorithm, Section~\ref{subsec:followups} identifies a follow-up strategy, and Section~\ref{subsec:upperlimits} explains how upper limits are computed from a non-detection.

\subsection{\fastt{} overview \label{subsec:fasttracks}}

\fastt{} \cite{Tenorio:2024jgc} provides a framework for evaluating short-coherence detection statistics in parallel. A template $\vec \lambda$ is evaluated by computing its frequency track $f(t; \vec \lambda)$, identifying the segment frequency bin crossed by the track at each timestamp, and performing the weighted sum of their corresponding statistics. This statistic evaluation can be vectorized over large batches of templates with \jax{}~\cite{jax2018github} and executed on both CPUs and GPUs. 

This framework has been applied to all-sky searches for sources in circular binary systems with data from the fourth~\citep{LIGOScientific:2025ouy} LVK observing run. The same approach is well suited to the directed search methodology presented here. For our targets, the sky position is known precisely and the size of the parameter space is instead driven by the unknown frequency and spindown together with the five orbital parameters (see Section~\ref{sec:template_density}).
Since \fastt{} makes no assumption on the functional form of $f(t;\vec \lambda)$, it can be applied to any CW search on SFT-like data, such as performing a search with the eccentric model of Equation~\ref{eq:ecc_freq_evolution}.

Following Ref.~\cite{Tenorio:2024jgc}, the SFTs are normalized by the 101-bin running-median estimate of the single-sided noise power spectral density $S_{\mathrm{n}}$ for each SFT and frequency. Identifying each SFT value at its starting time and frequency by $x(t;f)$, the normalized power is defined as 
\begin{equation}
    s(t;f) = \frac{4|\tilde{x}(t;f)|^2}
    {T_{\mathrm{SFT}} S_{\mathrm{n} X \alpha}(f)},
\end{equation}
where the indices $\alpha$ and $X$ label the SFT and the detector, respectively. 
For a template $\vec \lambda$, the weighted normalized power is 
\begin{equation}
    s(\vec \lambda) = \sum_{X,\alpha} w_{X\alpha}\,s\!\left(t_{X\alpha};\, f(t_{X\alpha};\vec \lambda)\right),
    \label{eq:weighted_power}
\end{equation}
where $X$ labels the detector and $\alpha$ the SFT. The weights account for both the noise level of each SFT and the antenna response of the detector,
\begin{equation}
    w_{X\alpha} \propto \frac{a^2(t_{X\alpha};\hat{n}) + b^2(t_{X\alpha};\hat{n})}{S_{\mathrm{n}X\alpha}},
    \qquad \sum_{X,\alpha} w_{X\alpha} = 1,
    \label{eq:weights}
\end{equation}
with $a$ and $b$ the amplitude-modulation coefficients \cite{Jaranowski:1998qm}, computed at the target's sky location.

We rank templates by the Critical Ratio (CR), with larger values indicating a greater excess above the expected noise level,
\begin{equation}
    \mathrm{CR}(\vec \lambda) = \frac{s(\vec \lambda) - \mu}{\sigma},
    \label{eq:critical_ratio}
\end{equation}
where $\mu$ and $\sigma$ are the mean and standard deviation of $s$, computed from the imported data.
This normalization makes CR values comparable across frequency bands.

\subsection{Follow-ups \label{subsec:followups}}

Interesting candidates are identified for a successive follow-up.
Usually, broad parameter-space searches, such as the one described in this work, select outliers for this additional step based on a threshold on the CR.
The threshold is related to the available computing power and dictates the sensitivity of the search~\cite{Mirasola:2024lcq, PhysRevD.97.103020, Brady:1997ji}.
Additional vetoes may be included to account for non-astrophysical features present in the data~\cite{O4LIGODetector:2026okh, Jaume:2024eob}.

During the follow-up, the coherence time of the search is progressively increased through multiple stages and the astrophysical origin of the candidate is assessed~\cite{PhysRevD.104.084012, Mirasola:2024lcq, PhysRevD.97.103020} by comparing the evolution of the detection statistic.

We use \pyfstat{}~\cite{pyfstat, Keitel:2021xeq} as a flexible follow-up pipeline suitable to tackle the challenges associated with the \gaia{} targets.
\pyfstat{} is based on a parallel-tempering ensemble Markov Chain Monte Carlo sampler as implemented in \ptemcee~\cite{Foreman-Mackey:2012any, Vousden:2016eeu}.
A new implementation~\cite{Ferrer-Martinez:2026lkg} based on bilby~\cite{bilby_paper} has also been recently made available to use additional sampling methods.
The pipeline evaluates the multi-detector $\F$-statistic~\cite{Jaranowski:1998qm, Cutler:2005hc} through LALSuite~\cite{lalsuite,swiglal} where the high-eccentricity binary Doppler correction is already implemented.
As reported in~\cite{PhysRevD.104.084012}, the output of each stage can be compared with the previous one(s) to distinguish noise-related from signal-related outliers.

\subsection{Upper limits \label{subsec:upperlimits}}

In the absence of a detection, we characterize the sensitivity of a search through simulated signals in the considered data set, following standard procedures in CW searches (e.g.,~\cite{LIGOScientific:2025ouy}). Simulated signals are added to the data and narrow ranges around them are analyzed with the same setup as the original search. Compared to all-sky searches for sources in circular binaries~\cite{LIGOScientific:2025ouy}, the higher dimensionality of our parameter space requires the recovery criterion to be tuned to the targeted systems.

We define a measure of closeness between a search template $\vec\lambda_{\mathrm{rec}}$ and the injected signal $\vec\lambda_{\mathrm{inj}}$ in units of the template spacing $\delta\lambda$ (Section~\ref{sec:metric}). A natural measure is the Euclidean distance~\cite{PhysRevD.110.123043}
\begin{equation}
    D_{\mathrm{E}} = \sqrt{\sum_{\lambda}\left(\frac{|\lambda_{\mathrm{inj}}-\lambda_{\mathrm{rec}}|}{\delta\lambda}\right)^{2}}\,.
    \label{eq:euclidean_distance}
\end{equation}
However, since we are studying a high-dimension parameter space, non-trivial correlations between parameters may exist, and the Euclidean distance may not be the most appropriate quantity for a detection criterion.

Therefore, we base our sensitivity estimates on a bin-wise detection criterion, where we require each parameter individually to lie within
$d_{\mathrm{th}}$ template spacings of the injection, namely 
\begin{equation}
    \max_{\lambda}\left(\frac{|\lambda_{\mathrm{inj}}-\lambda_{\mathrm{rec}}|}{\delta\lambda}\right) < d_{\mathrm{th}}\,.
    \label{eq:binwise_detection}
\end{equation}
Alternative closeness measures exist, such as the frequency distance of Ref.~\cite{Tenorio:2020cqm}. 

In addition, an injection is considered recovered if at least one template with $\mathrm{CR} \geq \mathrm{CR}_{\mathrm{th}}$ satisfies Equation~\ref{eq:binwise_detection}, where $\mathrm{CR}_{\mathrm{th}}$ is the threshold used to select candidates for follow-up (Section~\ref{subsec:followups}).

Upper limits ($h_{\rm UL}$) are computed by performing a set of injections at fixed amplitude $h_0$ and measuring the fraction of recovered injections, which defines the detection efficiency. Searches set $h_{\rm UL}$ as the amplitude at which a given detection efficiency is achieved, usually 95\% or 90\%. 
An equivalent way of expressing this upper limit is through the dimensionless sensitivity depth~\cite{Dreissigacker_2018, Behnke_2015, 2023APh...15302880W}
\begin{equation}
    \mathcal{D} \equiv \frac{\sqrt{S_n\cdot \rm Hz}}{h_{\rm UL}}\,.
    \label{eq:sensitivity_depth}
\end{equation}

%% file: implementation.tex
We can further improve the \fastt{} performance for these targets by taking advantage of their wide orbits.
The long $P_{\rm orb}$ of the binary systems (see Figure~\ref{fig:orbital_parameter_space}) implies that $\dot R/c$ slowly varies over time,
enabling us to downsample the time series and reduce the computational cost.
Additionally, the orbital parameters are known within a certain accuracy, allowing us to employ a dedicated template-placement strategy.
Here, we discuss the specific implementation designed for these targets in Section~\ref{subsec:Rdot} and in Section~\ref{subsec:batches}, which are dedicated to the $\dot R/c$ downsampling and to the template placing tailored for the \gaia{} DR3 candidates we are studying, respectively.

\subsection{Improving the binary Doppler calculation\label{subsec:Rdot}}

The most demanding part of the track evaluation is the binary Doppler modulation (Equation~\ref{eq:Rdot_over_c}) due to the transcendental nature of $E$ (Equation~\ref{eq:ecc_anomaly}). Since this term must be solved numerically for each template, it introduces a significant computational cost.

In SFT-based searches, the frequency track has to be evaluated for each SFT timestamp.
The targets we consider here are characterized by wide orbits, implying that $\dot{R}/c$ does not significantly change from one evaluation to the following one.
To speed up the computation, we can therefore evaluate $\dot{R}/c$ at a downsampled rate.
It can then be interpolated to the required resolution and used to compute the track.
This approach reduces the computational cost by decreasing the number of eccentric anomaly evaluations.
However, this may affect the CR of the template because of the introduced discrepancies between the interpolated and exact $\dot{R}/c$.

To quantify the aforementioned effect, we consider 5000~s-long SFTs covering the O4b~\citep{LIGOScientific:2026jgl} timestamps and perform tests for several sampling intervals, ranging from $10000\,\mathrm{s}$ to $100\,\mathrm{days}$ (corresponding to downsamplings between 0.5 and $\sim$0.0006).
Additionally, we study linear and cubic interpolation (limited by interpolation types compatible with \jax{}~\citep{Frostig2018,jax2018github,conlin_interpax_2026}).
The effect is quantified in terms of the maximum number of frequency bins ($N_{\rm{drift}}$) separating the interpolated and exact $\dot{R}/c$, computed as~\cite{PhysRevD.110.123043, Leaci:2016oja, Tenorio:2020cqm}
\begin{equation}
    N_{\rm{drift}} = \max{\left|\left(\frac{\dot{R}}{c}\right)_{\rm{interp}} - \left(\frac{\dot{R}}{c}\right)_{\rm{exact}}\right|\Tsft f_0}\,.
    \label{eq:rdot_interp_bin_drift}
\end{equation}

\begin{figure}[!t]
    \centering
    \includegraphics[width= \columnwidth]{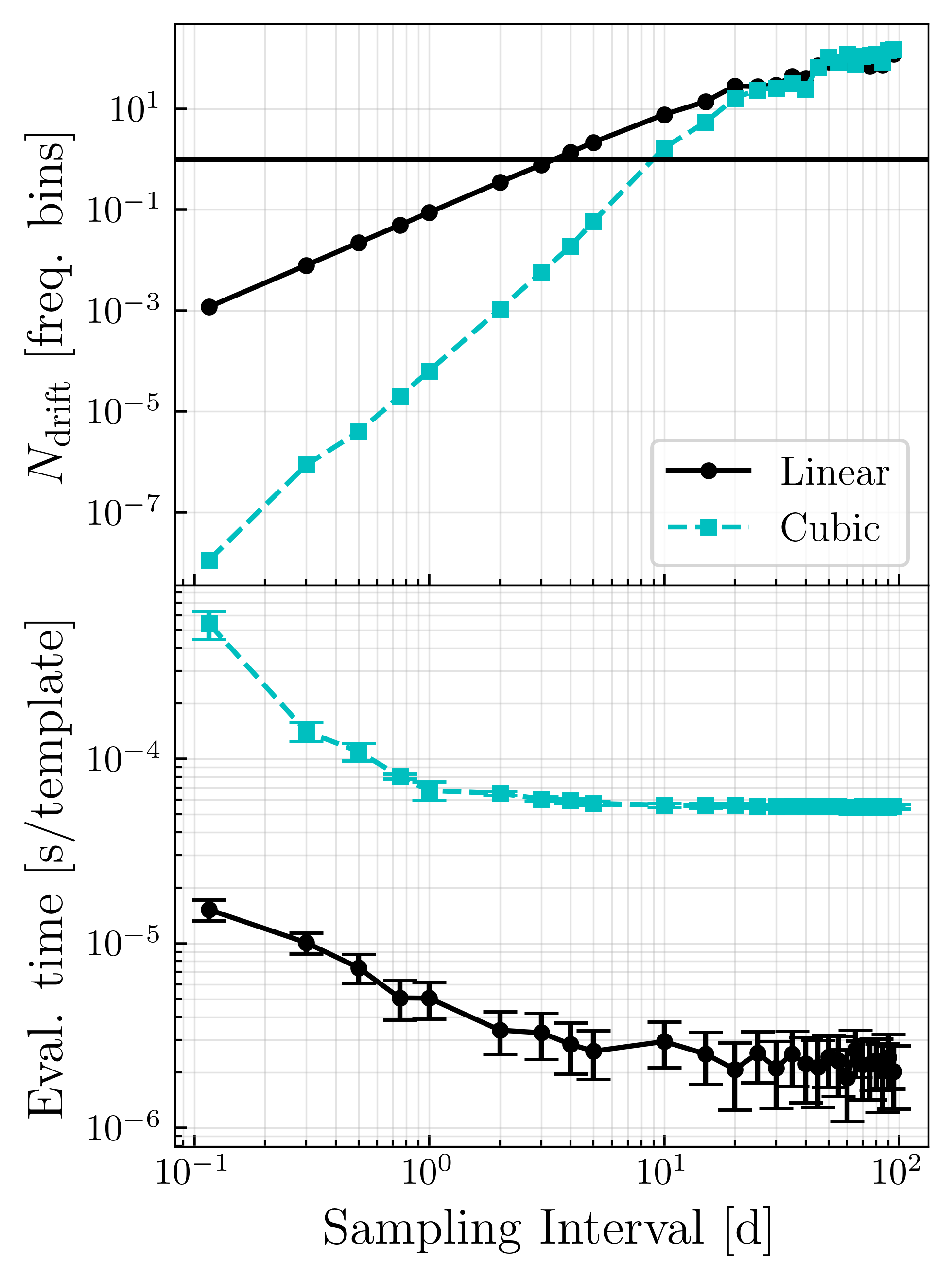}
    \caption{\textit{Upper panel}: Largest $N_{\rm{drift}}$ among the considered targets for different sampling intervals computed with Equation~\ref{eq:rdot_interp_bin_drift}. \textit{Lower panel}: Representative median per-template computational cost for the different parameters. See text for more details.}
    \label{fig:Rdot}
\end{figure}

For our tests, we explore a grid of all combinations of each orbital parameter $\lambda$ at $[\pm3,0]\sigma_{\lambda}$ using the \gaia{} uncertainties, which results in $3^5$ total templates.
Figure~\ref{fig:Rdot} shows the target with the largest $N_{\rm{drift}}$ for $f_0 = 200$~Hz across all configurations and targets considered in this study (\gaia{} DR3 ID 465093354131112960).
It also presents the median runtime needed to compute all the $\dot R/c$ for all the templates using \jax{} on an NVIDIA RTX 2070 laptop GPU.
We use the \texttt{Interpolator1D} routine in the \texttt{Interpax} package~\cite{conlin_interpax_2026} due to it supporting both linear and cubic interpolation.
The set of templates is re-computed $2000$ times for each spacing and interpolation technique, and the median per-template evaluation times with $1\sigma$ uncertainties are reported.

The error shown is the maximum error across all combinations of parameters.
As expected, increasing the sampling interval reduces the evaluation time per template, while $N_{\rm{drift}}$ grows monotonically.
For both interpolation methods, we observe a plateau in computing cost for intervals greater than about 2~days, implying diminishing returns above that value.
The cubic interpolation achieves better performances in terms of approximation, but corresponds to a computing cost systematically higher than the linear interpolation by an order of magnitude across the entire studied range.

Similarly to other search methodologies, we can set a maximum tolerance of $N_{\rm{drift}}=1$ (see, e.g.,~\cite{Astone:2000jza, Astone:2014esa}).
However, we adopt a conservative choice of substantially lower values to ensure that interpolation errors remain negligible in this exploratory study.
Therefore, we select linear interpolation with a sampling interval of $0.5$~days.
This represents a tradeoff between computing cost and accuracy.
After making this decision, we use the \texttt{jax.numpy.interp} routine natively built into \jax{}~\citep{Frostig2018,jax2018github} for compatibility with the latest release (v0.10.2~\cite{jax2018github_02} at the time); this does not affect the conclusions reported in the text as we have found that the two methods perform comparably in our testing environment.

\subsection{Template Placement\label{subsec:batches}}

The \gaia{} DR3 NS candidates that motivated this study (see Section~\ref{sec:search_parameter_space}) have estimates for the binary orbital parameters but no information about their $f_0$ and $f_1$.
However, as shown in Section~\ref{sec:metric}, the uncertainties on the orbital parameters are far from negligible.
Compared to previous \fastt{} all-sky searches, where all the parameters are unknown, we can use \gaia{}'s observations to tailor our template placement.

We implement a hybrid approach, where we construct a grid in the $f_0-f_1$ space using Equations~\ref{eq:delta_f0} and~\ref{eq:delta_f1}, while the orbital parameters are randomly sampled.
For each $f_0$ and $f_1$ combination, we draw values of binary parameters from Gaussian distributions centered at the \gaia{} DR3 fitted values with standard deviations set to their uncertainties.
In this process we account for the covariance between $P_{\mathrm{orb}}$ and $t_p$ that arises from propagating $t_p$ forward in time via a covariance matrix $\boldsymbol{\Sigma}_{\rm{DR3}}$ by sampling from a multivariate Gaussian distribution for these two parameters $\mathcal{N}_2(\boldsymbol{\mu}_{\rm{DR3}},\boldsymbol{\Sigma}_{\rm{DR3}})$~\citep{Leaci:2015bka,LIGOScientific:2026ppd} with $\boldsymbol{\mu}_{\rm{DR3}}=(\mu_{t_p}, \mu_{P_{\rm orb}})$ and covariance matrix defined as
\begin{equation}
    \boldsymbol{\Sigma}_{\rm{DR3}} = 
    \begin{pmatrix}
        \sigma_{t_p}^2 + N_{\rm orb}^2\sigma_{P_{\rm orb}}^2 & N_{\rm orb}\sigma_{P_{\rm orb}}^2 \\
        N_{\rm orb}\sigma_{P_{\rm orb}}^2 & \sigma_{P_{\rm orb}}^2
    \end{pmatrix}\,,
    \label{eq:2d_norm}
\end{equation}
where $N_{\rm orb}$ being the number of orbits from the initially-reported epoch to O4b.

We generate $N_{\rm tot}$ binary parameter combinations following the discussion in Section~\ref{sec:metric} from these distributions.
The number of binary templates is computed for a parameter space of $\pm3\sigma$ in each parameter using Equation~\ref{eq:delta_orbital} to effectively explore the tails of the $1\sigma$ distributions.
For a given $f_0$, we use the same set of binary orbital parameters across the entire $f_1$ range.
To save storage, we retain only the template with the loudest CR, computed through Equation~\ref{eq:critical_ratio}, for each unique $f_0-f_1$ combination, essentially maximizing over all binary orbital parameters.

%% file: injections.tex
To demonstrate the effectiveness of our pipeline, we perform injections into both H1 and L1 simulated Gaussian noise and real O4b data (see Section~\ref{sec:data} for more information).
First, Section~\ref{sec:injection_camp} introduces the parameter-space region covered by our tests.
Then, Section~\ref{sec:gauss_noise_inj} shows the performance of the method in Gaussian-noise data, while Section~\ref{sec:real_noise_inj} focuses on injections in real data.

\subsection{Injection campaign\label{sec:injection_camp}}

Our injection campaign focuses on two NS candidates from Section~\ref{sec:search_parameter_space}: 1350295047363872512, 4922744974687373440.
According to the estimation in Section~\ref{sec:template_density}, they are the cheapest and a reasonably-expensive candidates, respectively.
The conclusions we draw from them are valid for the entire set of targets, with differences related only to trials factors (see Figure~\ref{fig:tot_num_templates}).

Given the frequency range of interest for the \gaia{} DR3 NS candidates, we study the performances of our pipeline in the ``low''- and ``high''-frequency regimes.
This allows us to study the impact of the frequency-dependent trials factor shown in Figure~\ref{fig:tot_num_templates}, which can significantly affect the sensitivity of our proposed search.
More specifically, we simulate signals drawing from uniform distributions within $51.1-51.3\,\rm{Hz}$ and $198.1-198.3\,\rm{Hz}$.

Following the discussion in Section~\ref{sec:search_parameter_space}, the CW frequency evolution of the \gaia{} NS candidates is likely dominated by their intrinsic spin-down.
Therefore, we account for this by including a random $f_1\in[-10^{-9}, 0]$~Hz/s (e.g.,~\cite{LIGOScientific:2025ouy}), drawn from a uniform distribution.

Each injection simulates a different sky location with $\delta$ fixed at the \gaia{} measured value, while $\alpha$ is randomly sampled from a uniform distribution that excludes a region of $\pm\pi/4$ around the target's one.
This ensures that we would not be simulating signals on top of a potentially real one, while maintaining approximately equal antenna response among all the considered sky locations, as the $\alpha$ dependence averages out over months of data.

For the binary orbital parameters, we pick values in the same way as the template placement procedure described in Section~\ref{subsec:batches}.

Lastly, we inject signals with $\cos\iota\in[-1,1]$, $\psi\in[0,\pi/2]$ and $\phi_0\in[0,2\pi]$ (i.e., the phase of the signal at $t_{\rm ref}$), all drawn from uniform distributions.

We study a set of injected $\depth\in[30,40,50,70,90,110]$ to give an indicative estimation of the method's sensitivity.
For each $\depth$, we perform 100 injections sampling from the aforementioned distributions and run the search algorithm as described throughout this paper: we compute the total number of templates at a certain frequency following Section~\ref{sec:template_density} and place them according to Section~\ref{subsec:batches}; for each template, we compute the track as in Section~\ref{subsec:Rdot} and evaluate the CR through Equation~\ref{eq:critical_ratio}.

All the injection parameters are summarised in Table~\ref{tab:inj_distributions}.

\begin{table}[!ht]
    \centering
    \renewcommand*{\arraystretch}{1.4}
    \begin{tabular}{||l | c||} 
        \hline
        Property & Value \\ 
        \hline\hline
        $f_0$ & [$\mathcal{U}(51.1-51.3\,\rm{Hz})$, $\mathcal{U}(198.1-198.3\,\rm{Hz})$] \\ 
        \hline
        $f_1$ & $\mathcal{U}(-10^{-9},0)$ \\
        \hline
        $\alpha$ & $\mathcal{U}(0,2\pi)\notin(\alpha_{\rm{DR3}}-\pi/4,\alpha_{\rm{DR3}}+\pi/4)$ \\
        \hline
        $\delta$ & $\delta_{\rm{DR3}}$ \\
        \hline
        $a_p,e,\omega$ & $\mathcal{N}(\mu_{\rm{DR3}},\sigma_{\rm{DR3}})$ \\
        \hline
        $P_{\rm orb},t_p$ & $\mathcal{N}_2(\boldsymbol{\mu}_{\rm{DR3}},\boldsymbol{\Sigma}_{\rm{DR3}})$\\
        \hline
        $\cos\iota$ & $\mathcal{U}(-1,1)$ \\
        \hline
        $\psi$ & $\mathcal{U}(0,\pi/2)$ \\
        \hline
        $\phi_0$ & $\mathcal{U}(0,2\pi)$ \\
        \hline
        $\mathcal{D}$ & [30,40,50,70,90,110] \\
        \hline
        
    \end{tabular}
    \caption{Parameter-space region covered by the injection campaign. Parameters of the form $\lambda_{\rm{DR3}}$ and $\Delta \lambda_{\rm{DR3}}$ are the \gaia DR3 reference values and uncertainties. $\mathcal{U}$ and $\mathcal{N}$ represent uniform and normal distributions, respectively. Injections were performed at multiple discrete sensitivity depths $\mathcal{D}$. All parameters are drawn from uncorrelated distributions but $P_{\rm orb},t_p$, which are sampled from a bivariate normal related to the propagation of $t_p$ to the O4b epoch~\citep{Leaci:2015bka,LIGOScientific:2026ppd}.}
    \label{tab:inj_distributions}
\end{table}

To study the sensitivity of the search pipeline, we follow the discussion in Section~\ref{subsec:upperlimits}.
Therefore, we use the detection criterion in Equation~\ref{eq:binwise_detection} setting $d_{\mathrm{th}} = 2$, and explore a set of $\mathrm{CR}_{\mathrm{th}}\in[5,5.5,6,6.5,7]$.

The discussion and plots in this section focus on the low-frequency regime. Results for the high-frequency band are shown in Appendix~\ref{app:high_ecc_inj}.

\subsection{Gaussian noise injections\label{sec:gauss_noise_inj}}

We begin with injections into Gaussian noise simulated with $\sqrt{\Sn}= 10^{-23}\,\rm{Hz}^{-1/2}$. This demonstrates the performance of the analysis under ideal circumstances. The Gaussian noise was generated using the H1 and L1 timestamps of O4b~\cite{LIGOScientific:2026jgl, Goetz:T2400058} such that it matched the total duration and detector duty cycles, including gaps in observations.

\begin{figure*}[!ht]
    \includegraphics[width=0.45 \linewidth]{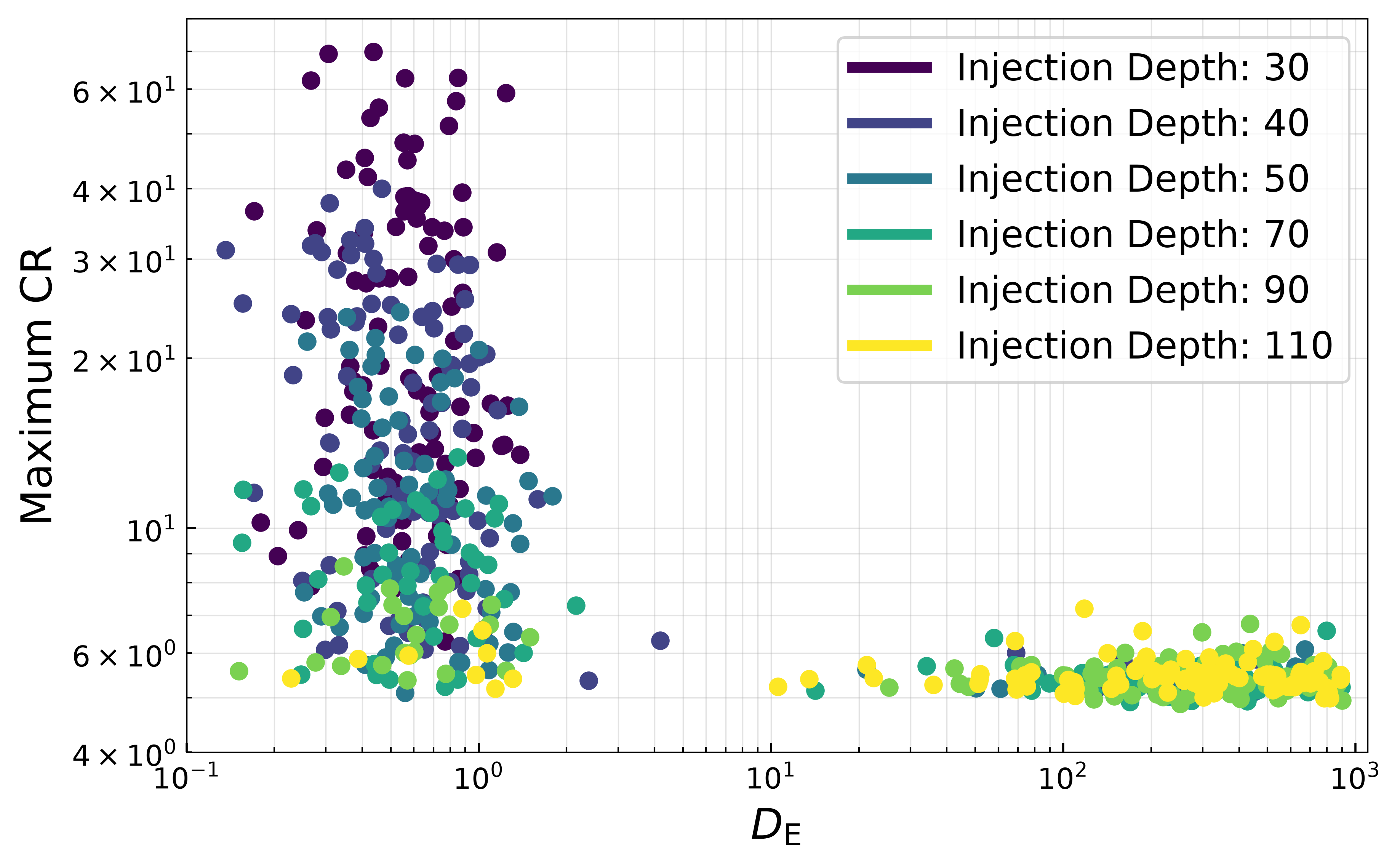}
    \includegraphics[width=0.45 \linewidth]{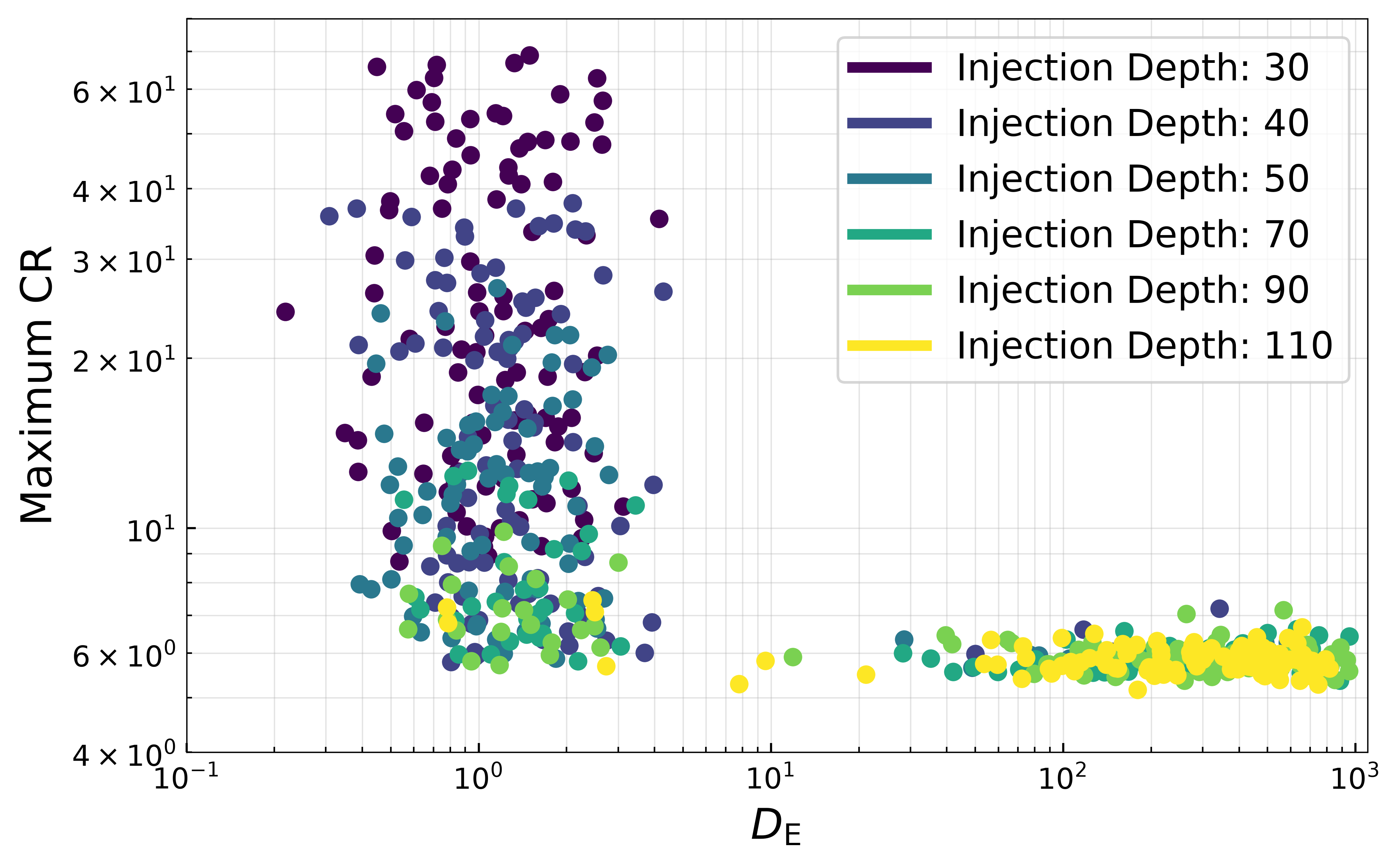}
    \includegraphics[width = 0.45\linewidth]{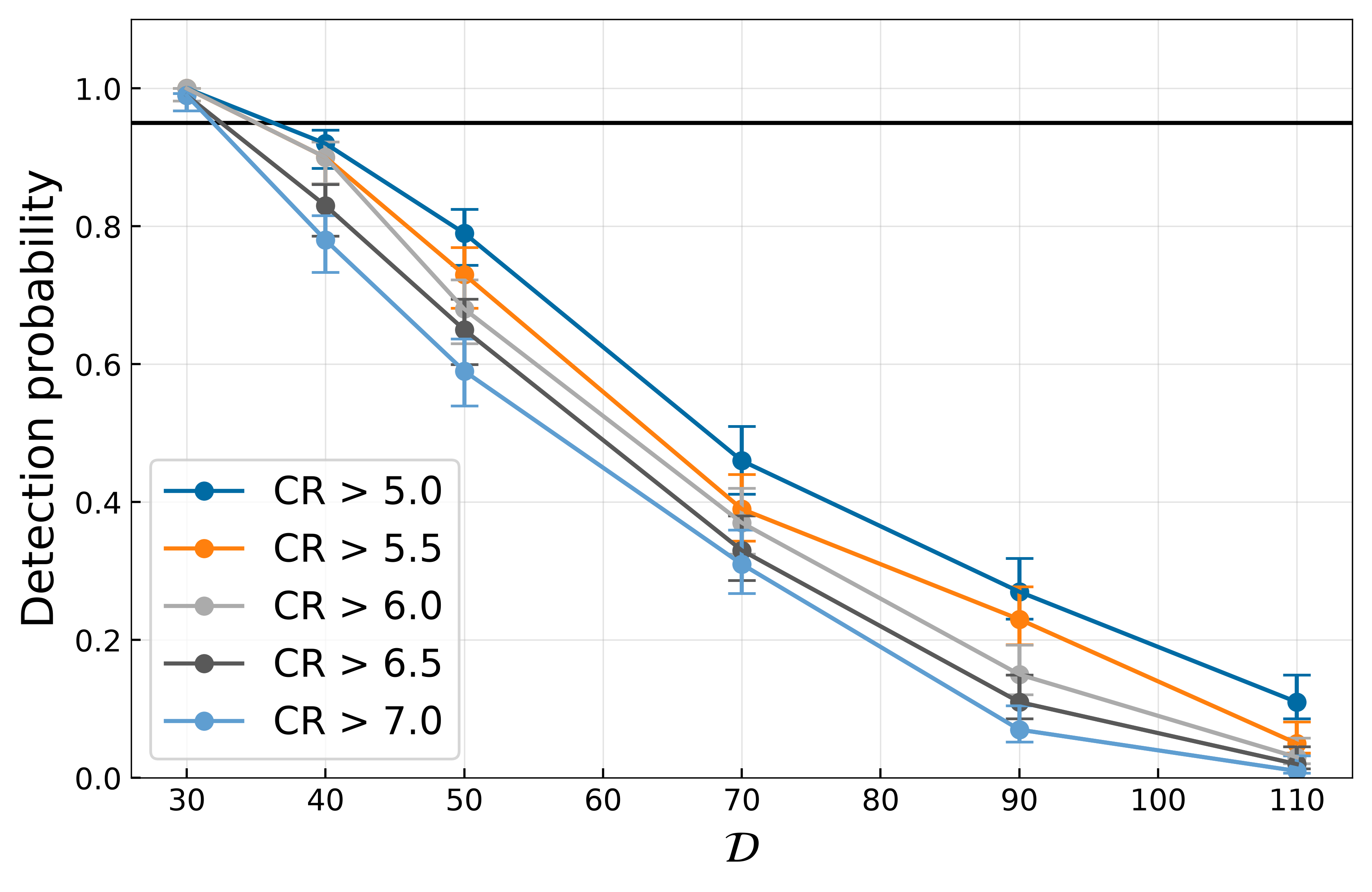}
    \includegraphics[width = 0.45\linewidth]{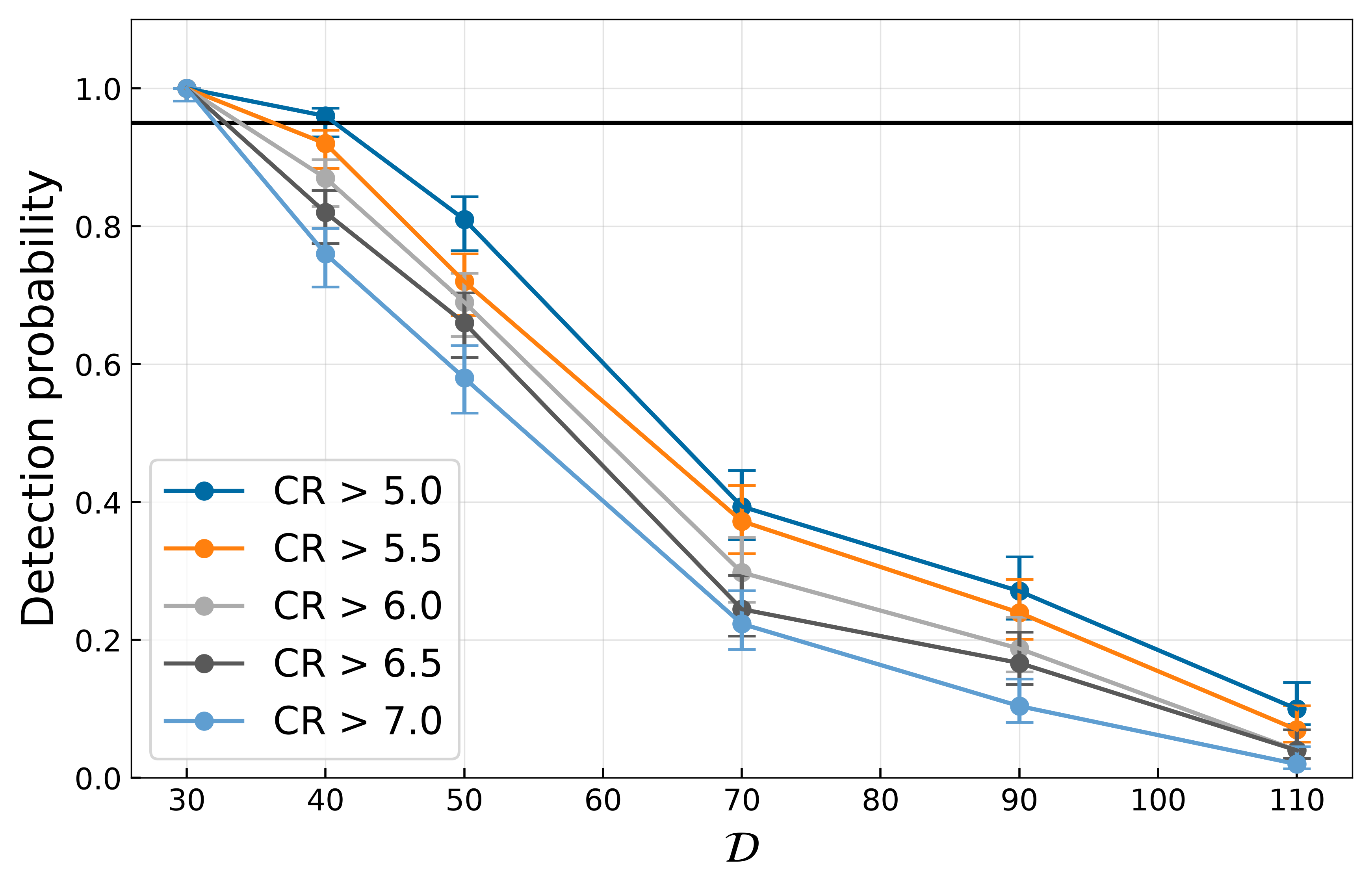}
    \caption{\textit{Upper panels}: Maximum CR recovered per injection plotted against the template's $D_{\mathrm{E}}$ (Equation~\ref{eq:euclidean_distance}) using simulated Gaussian noise at different sensitivity depths for the (\textit{left}) cheapest candidate and (\textit{right}) a more expensive one. Plots show injections within $51.1\,\mathrm{Hz}$ and $51.3\,\mathrm{Hz}$. \textit{Lower panels}: Detection probability as a function of injection depth for multiple $\rm CR_{\rm th}$ for the injection set presented in the upper panels. See Section~\ref{sec:injection_camp} for more details about these injections and Section~\ref{sec:gauss_noise_inj} for a discussion of these results.}
    \label{fig:lowf_injections_gaussian}
\end{figure*}

The upper panels of Figure~\ref{fig:lowf_injections_gaussian} show the distribution of $D_E$ (Equation~\ref{eq:euclidean_distance}) and the recovered maximum CR for each injection for the two studied \gaia{} DR3 candidates.
Our results match the expectations.
Both plots show the same features: a cluster at low $D_E$ for a broad range of CR and another at high $D_E$ with low CR.
The former corresponds to templates very close to the injections and likely to be detected if a low-enough $\rm CR_{\rm th}$ is used.
Within this cluster, we observe higher CRs for lower depths (i.e., louder signals) with a spread related to the $\cos\iota$ dependence.
Comparing the two searches, we observe that this cluster is centered at lower $D_E$ values for the cheapest one compared to the other, reflecting the increased trials factor for the more expensive target. 
The second cluster corresponds to noise-related templates with an average value around $\rm CR\sim 5.5$ for the cheapest target and $\rm CR\sim 6$ for the other.
This difference is, again, related to trials factors.

The lower panels of Figure~\ref{fig:lowf_injections_gaussian} show the detection probability as a function of $\depth$ computed as the fraction of injections detected.
As expected, the probability decreases with increasing depth with similar results for the two considered NS candidates.

\subsection{Real data injections\label{sec:real_noise_inj}}

\begin{figure*}
    \includegraphics[width=0.45 \linewidth]{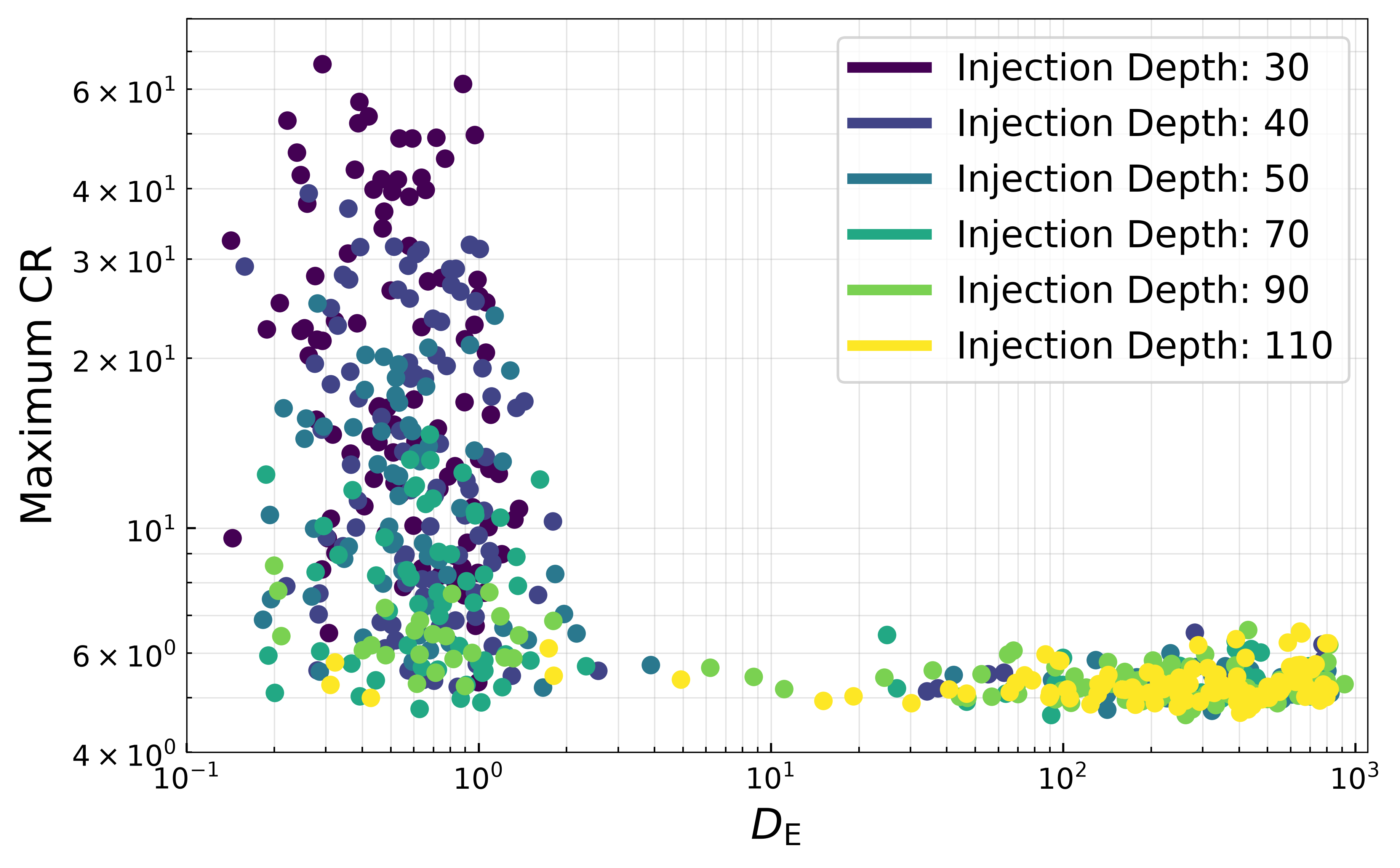}
    \includegraphics[width=0.45 \linewidth]{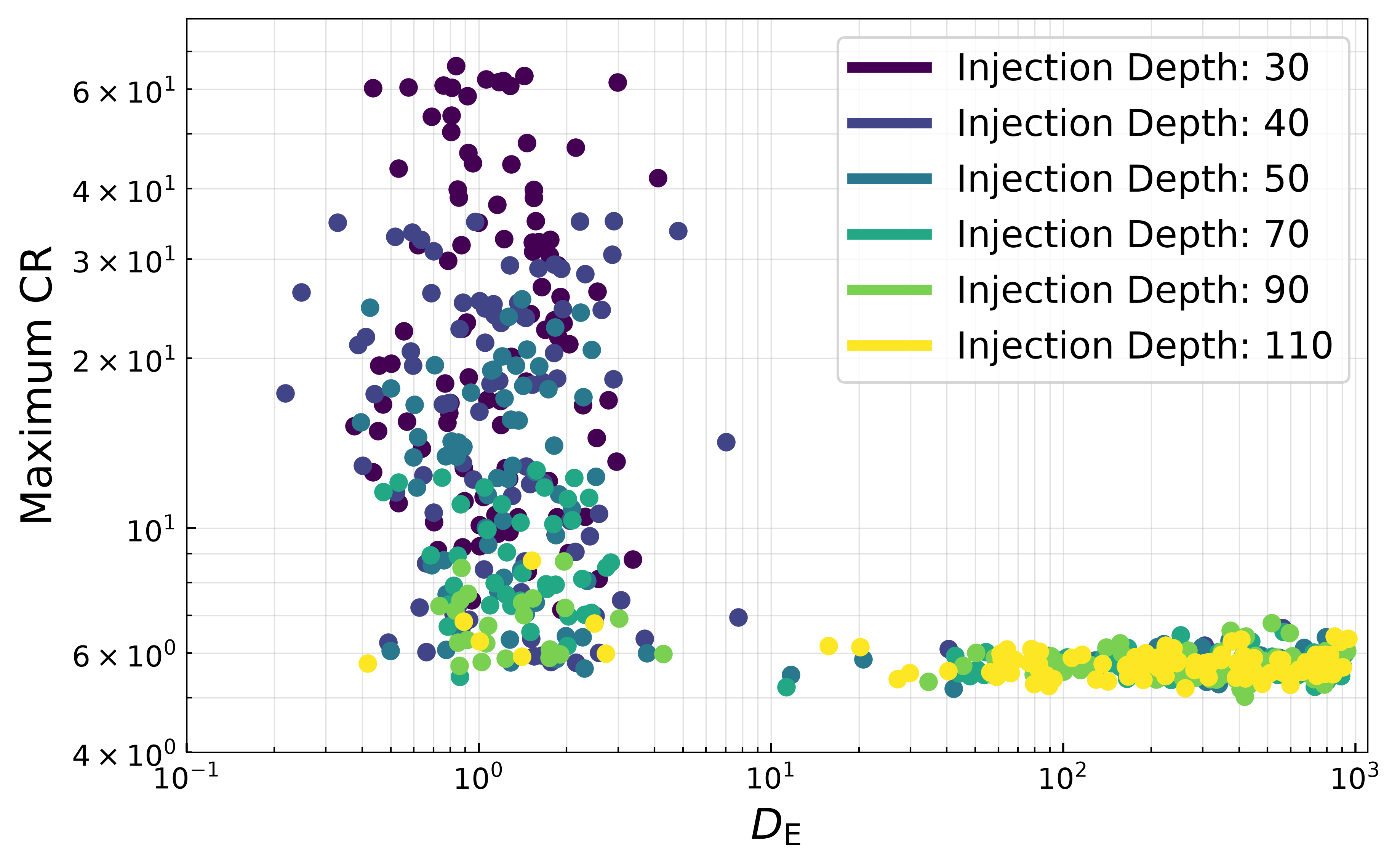}
    \includegraphics[width = 0.45\linewidth]{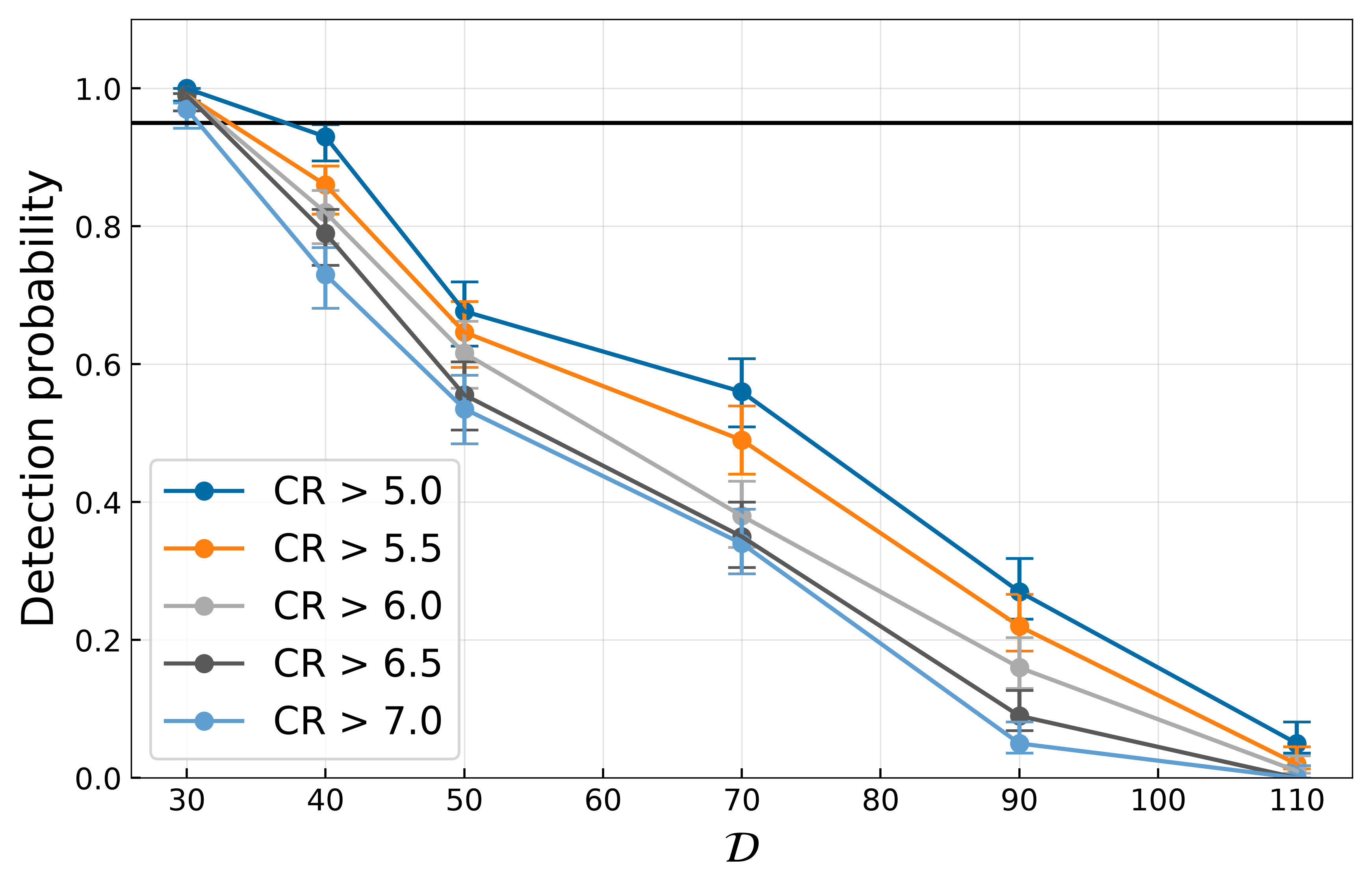}
    \includegraphics[width = 0.45\linewidth]{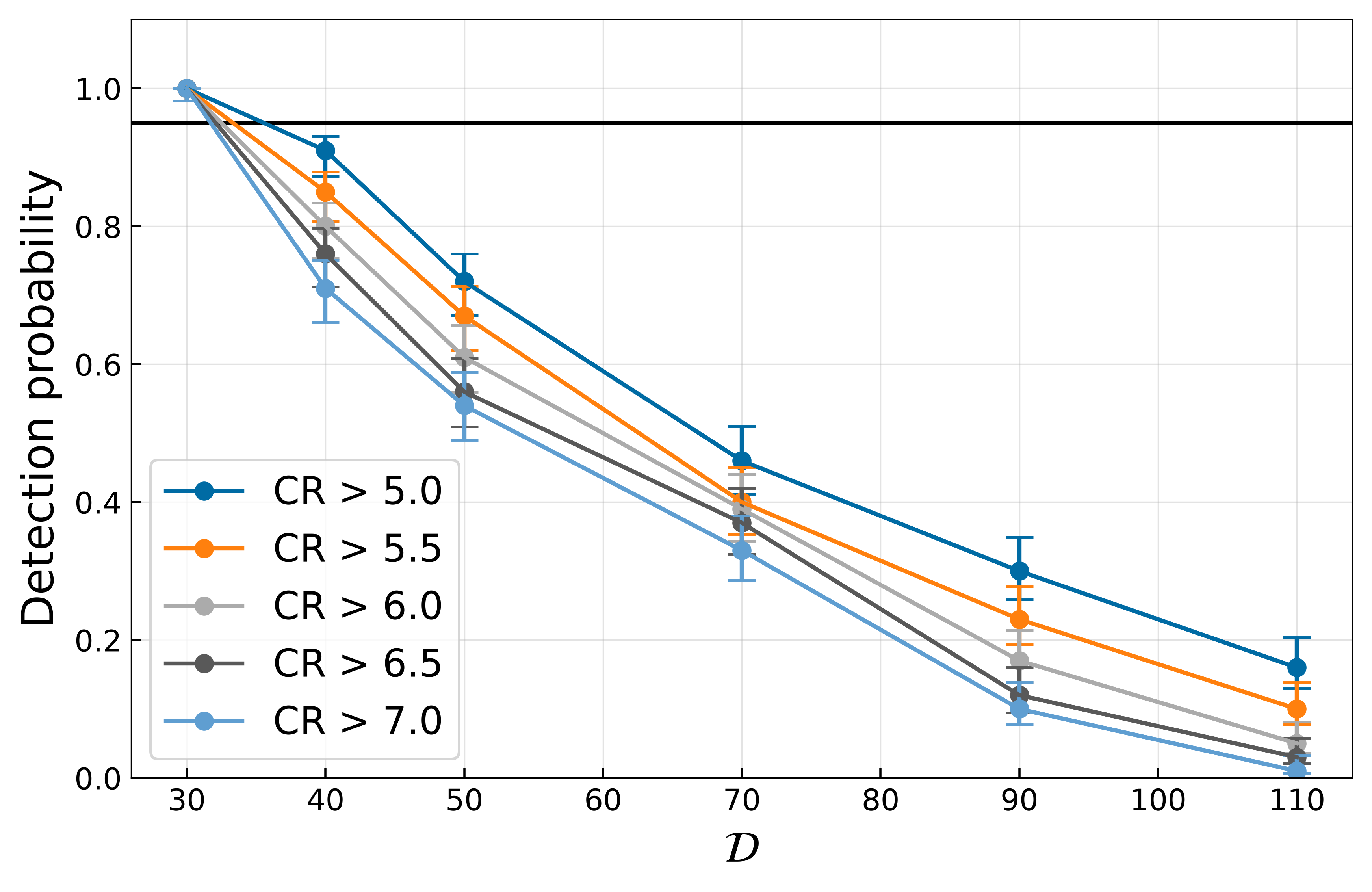}
    \caption{Same as Figure~\ref{fig:lowf_injections_gaussian} but for software injections into real detector data.}
    \label{fig:lowf_injections_real}
\end{figure*}

We perform additional injections into the real detector data from O4b using H1 and L1 data, pre-processed following the procedure detailed in Section~\ref{sec:data}.
We repeat the same analysis as described in Section~\ref{sec:gauss_noise_inj} for the same two targets.
Results are reported in Figure~\ref{fig:lowf_injections_real} in a similar way as before.
The outcome of this test is consistent with the previous one with conclusions virtually identical.
These results therefore suggest that the method retains comparable performance under realistic detector conditions to that obtained in the idealised stationary Gaussian-noise scenario reported in Section~\ref{sec:gauss_noise_inj}.

%% file: conclusion.tex
We have presented a new directed method to search for CWs from rotating NSs in wide highly-eccentric binary orbits.
This exploratory study is motivated by the 21 \gaia{} DR3 NS candidates reported in Ref.~\citep{2024OJAp....7E..58E}.
For all of them, we have an estimation of the binary orbital parameters, distance and sky location. 
However, no pulsation is observed and therefore their rotational frequencies and spin downs are unknown.

In the highly-eccentric regime that these \gaia{} targets live in, no full parameter-space metric has been derived yet to place templates and estimate the computing cost of an analysis covering a certain parameter-space region.
We tackle these problems by introducing a ``frequency metric'', following a similar criterion developed in Ref.~\cite{Tenorio:2024jgc} for \fastt{}.

The computation of the binary Doppler correction is another limiting factor for CW searches in the eccentric regime. The increased computational cost of the numerical solution is offset by computing the binary Doppler shift correction at a reduced resolution and interpolating to the required one. We have demonstrated that this downsampling and interpolation step can be adjusted to balance computational speed and interpolation accuracy.

Additionally, we present a novel approach to incorporating \gaia{} observations into CW searches through a hybrid template-placement strategy. While the frequency parameters are placed according to the frequency metric, the binary orbital parameters are sampled from distributions informed by the \gaia{} measurements and their uncertainties. This allows us to exploit the available electromagnetic information to optimize the exploration of the parameter space.

In order to demonstrate the effectiveness of this pipeline, we have performed mock searches for injected signals using both Gaussian noise and real detector data from the LVK O4b observing run.
The pipeline described here uses the \fastt{} infrastructure to implement the highly-eccentric binary orbital motion of the \gaia{} targets.
These injections demonstrate that such a search is feasible with real astrophysical targets.

The effort detailed in this work builds the foundations for a first of its kind CW search informed by \gaia{} observations.
Future observations of targets like the ones discussed here may improve the accuracy on the orbital parameters, significantly increasing the viability of CW searches targeting them.

%% file: appendix_derivatives.tex
The binary Doppler shift $\dot R/c$, Equation~\ref{eq:Rdot_over_c}, depends on the orbital parameters $\lambda\in\{a_p,\Omega,t_p,e,\omega\}$ both explicitly and implicitly through the eccentric anomaly $E(t)$.
Its derivatives therefore follow from the chain rule:
\begin{equation}
    \frac{\partial(\dot R/c)}{\partial\lambda}
    =
    \left.\frac{\partial(\dot R/c)}{\partial\lambda}\right|_E
    +
    \frac{\partial(\dot R/c)}{\partial E}
    \frac{\partial E}{\partial\lambda}\,.
    \label{eq:chain_rule}
\end{equation}
Implicit differentiation of Kepler's equation at fixed $t$
then yields
\begin{align}
    \frac{\partial(\dot R/c)}{\partial E} &= -\frac{a_p\Omega\left[\sqrt{1-e^2}\cos\omega\sin E + (\cos E-e)\sin\omega\right]}{(1-e\cos E)^2}\,, \\
    \frac{\partial(\dot R/c)}{\partial a_p} &= \frac{\dot R/c}{a_p}\,, \\
    \frac{\partial(\dot R/c)}{\partial\omega} &= \frac{a_p\Omega\left[-\cos\omega\sin E - \sqrt{1-e^2}\cos E\sin\omega\right]}{1-e\cos E}\,, \\
    \frac{\partial(\dot R/c)}{\partial t_p} &= -\frac{\Omega}{1-e\cos E}\,\frac{\partial(\dot R/c)}{\partial E}\,, \\
    \frac{\partial(\dot R/c)}{\partial\Omega} &= \frac{\dot R/c}{\Omega} + \frac{t-t_p}{1-e\cos E}\,\frac{\partial(\dot R/c)}{\partial E}\,, \\
    \frac{\partial(\dot R/c)}{\partial e} &= \frac{a_p\Omega\cos E}{(1-e\cos E)^2}\left[\frac{\cos E-e}{\sqrt{1-e^2}}\cos\omega - \sin\omega\sin E\right] \notag\\
    &\quad + \frac{\sin E}{1-e\cos E}\,\frac{\partial(\dot R/c)}{\partial E}\,.
    \label{eq:orbital_derivatives}
\end{align}

%% file: appendix_addtl_injections.tex
Here, we present additional results for injections at the high end of the frequency range we considered (between $198.1\,\rm{Hz}$ and $198.3\,\rm{Hz}$). These are for the same \gaia{} target as the injections performed in Section~\ref{sec:injections}.

\begin{figure*}[!ht]
    \includegraphics[width=0.45 \linewidth]{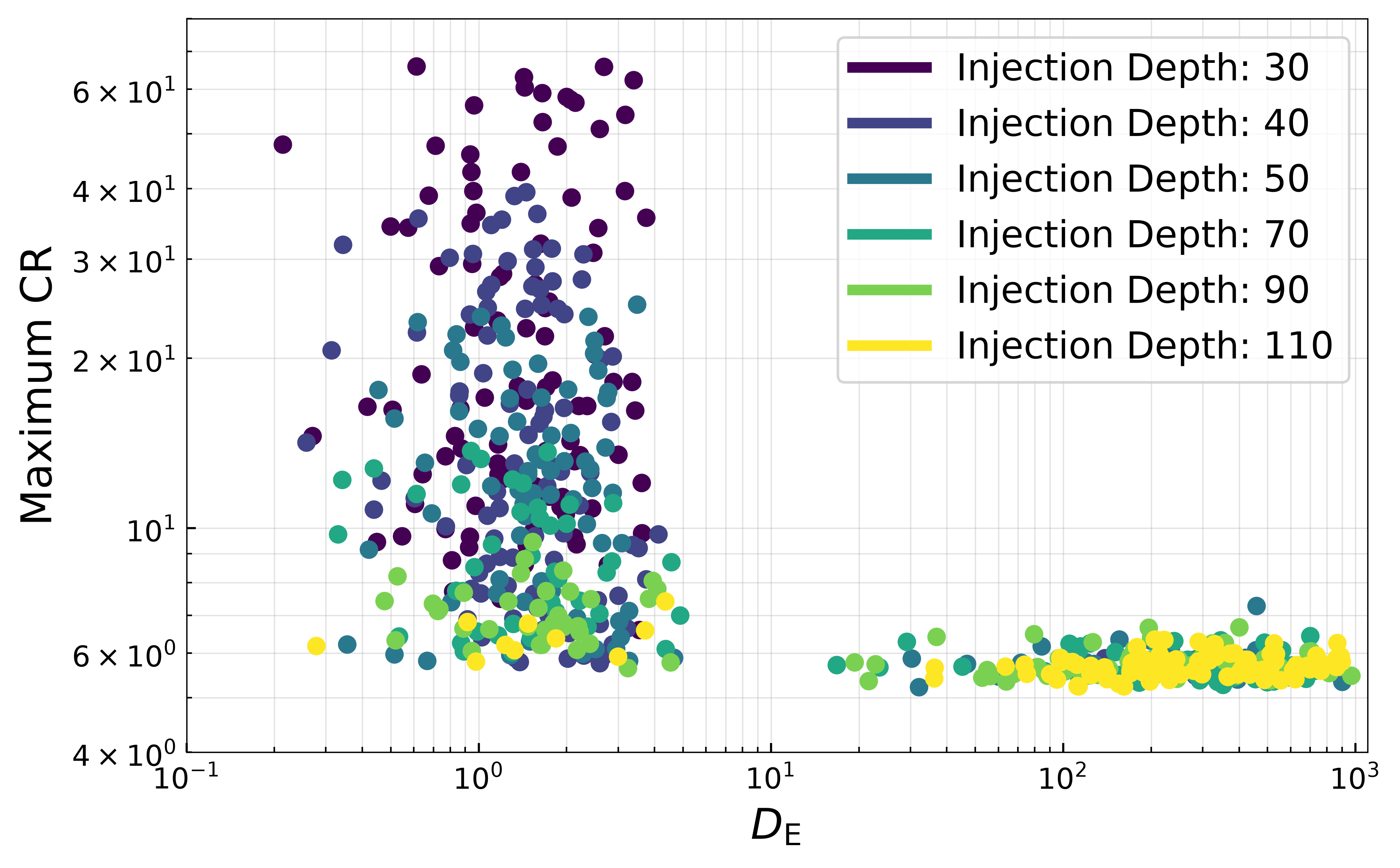}
    \includegraphics[width=0.45 \linewidth]{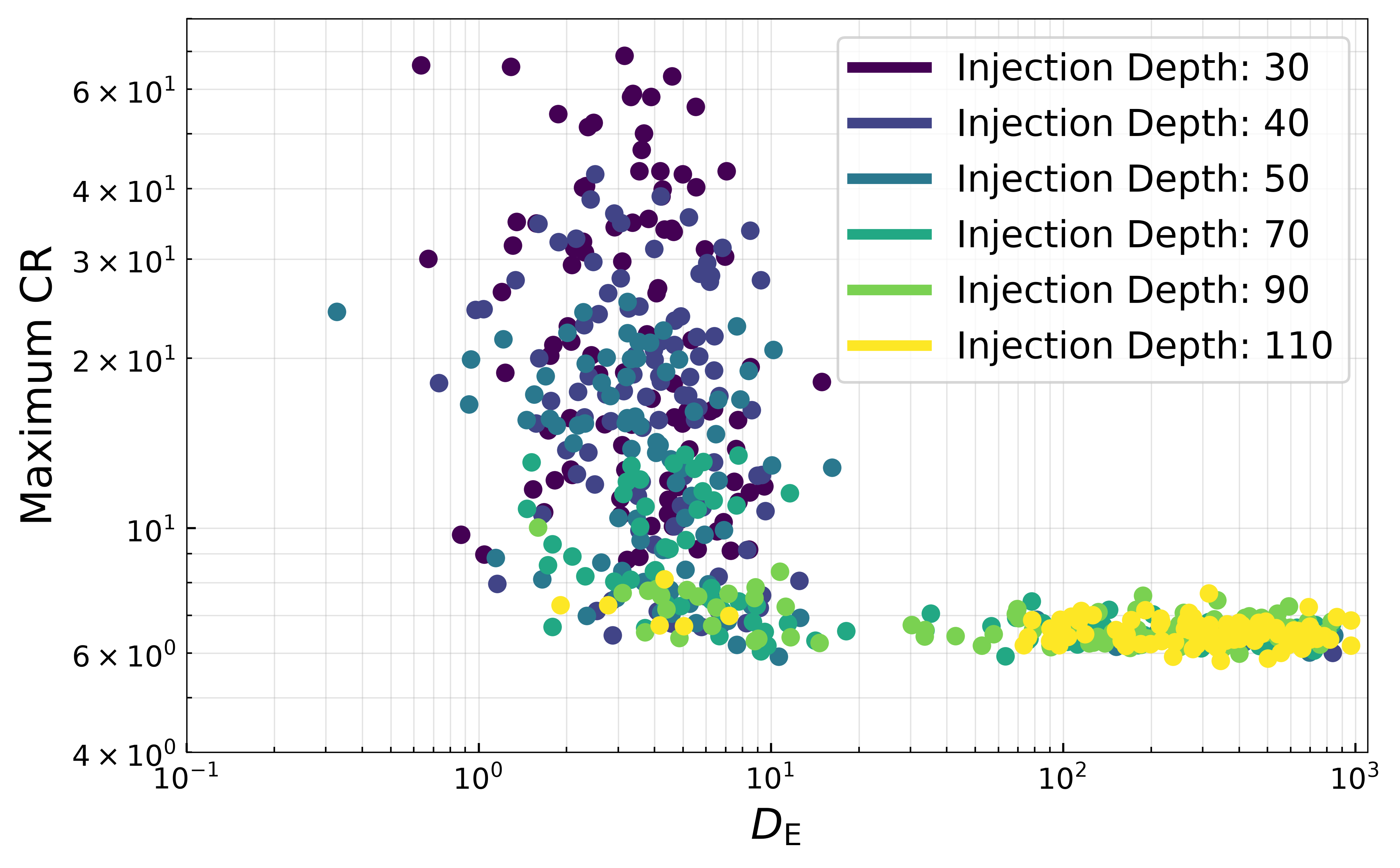}
    \includegraphics[width = 0.45\linewidth]{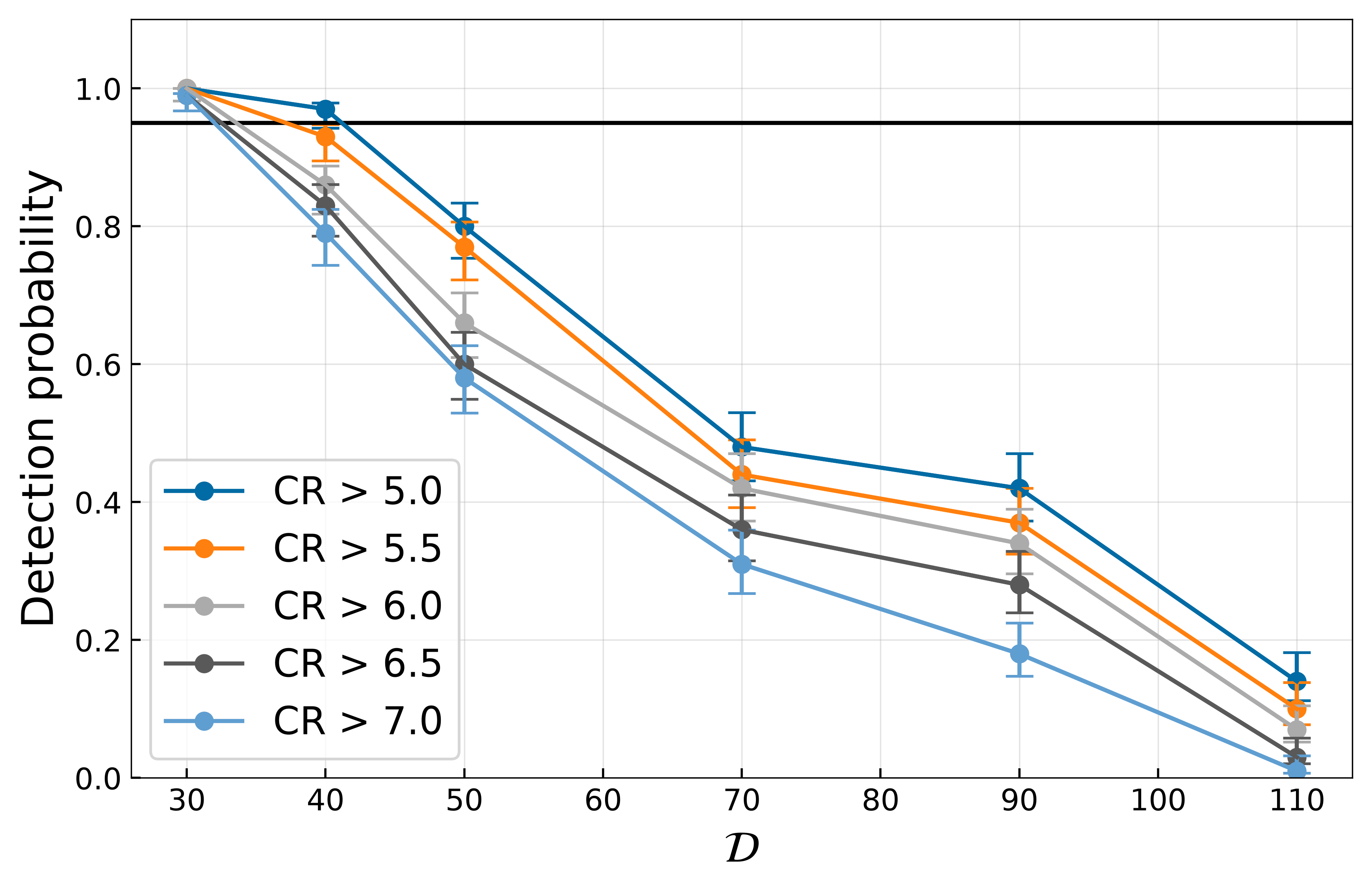}
    \includegraphics[width = 0.45\linewidth]{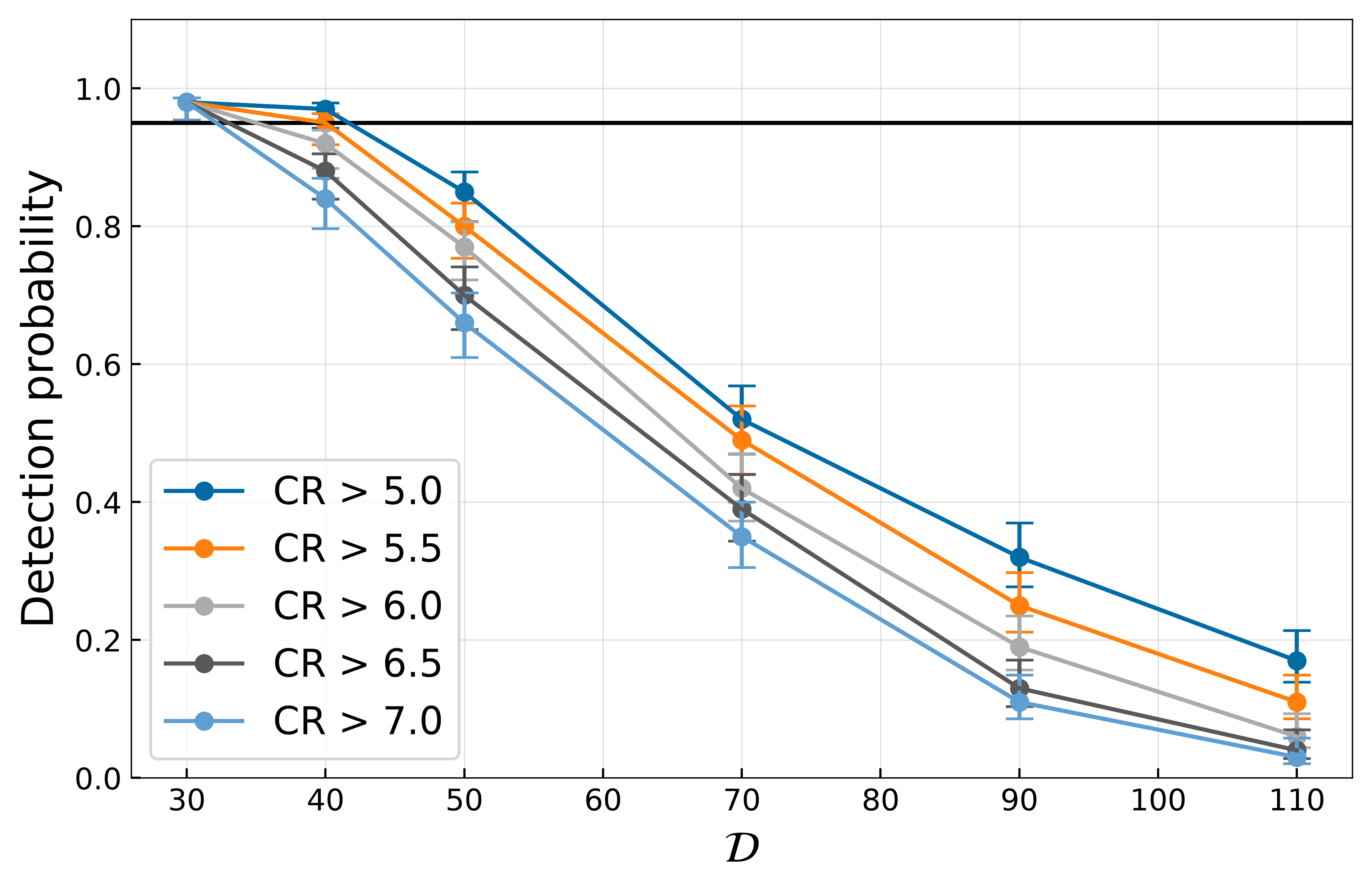}
    \caption{Same as Figure~\ref{fig:lowf_injections_gaussian} but for injections with $f_0$ between $198.1\,\mathrm{Hz}$ and $198.3\,\mathrm{Hz}$.}
    \label{fig:highf_injections_gaussian}
\end{figure*}

Figure~\ref{fig:highf_injections_gaussian} shows injections into simulated Gaussian noise for the same \gaia{} target as in Section~\ref{sec:gauss_noise_inj}.

\begin{figure*}[!ht]
    \includegraphics[width=0.45 \linewidth]{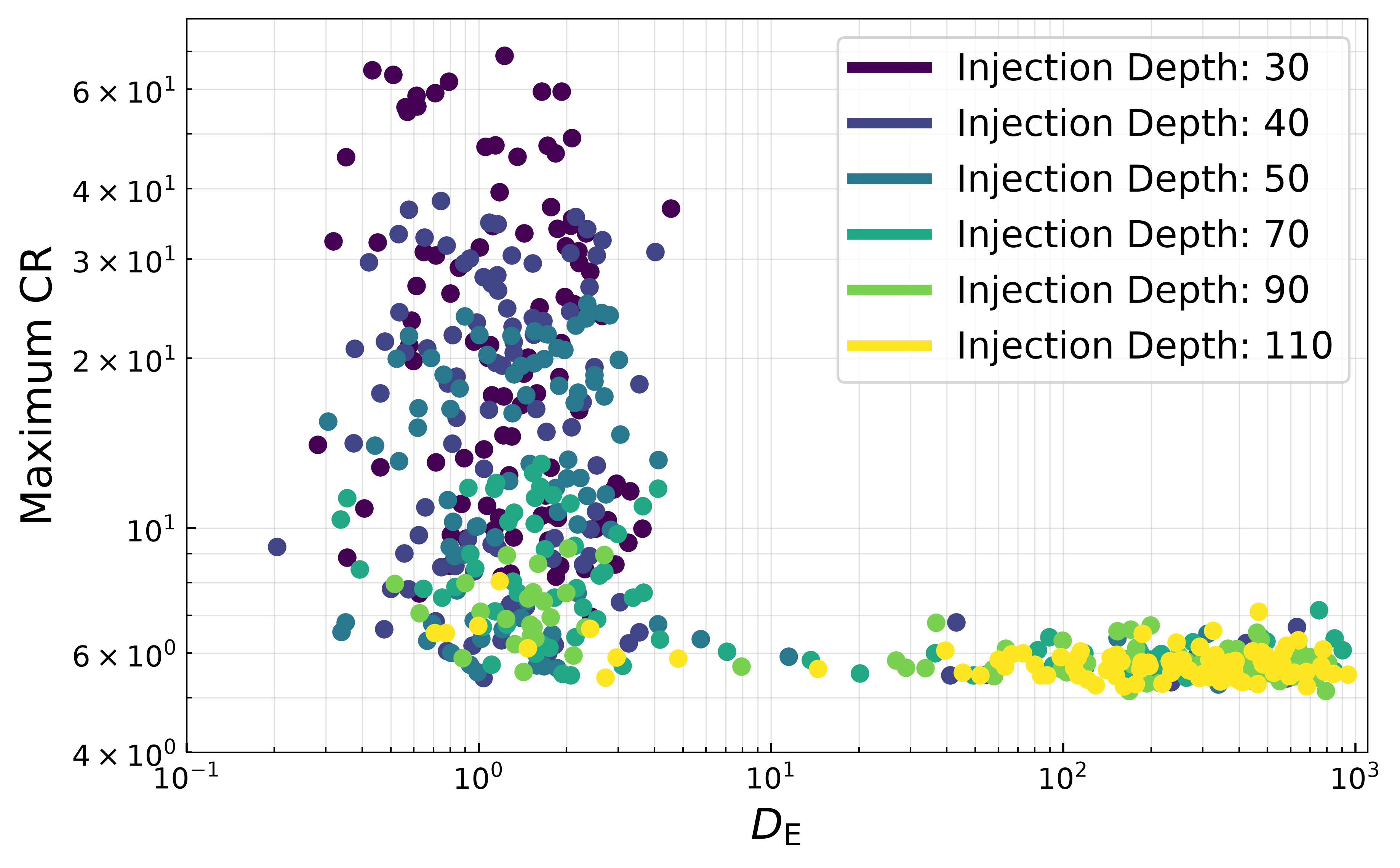}
    \includegraphics[width=0.45 \linewidth]{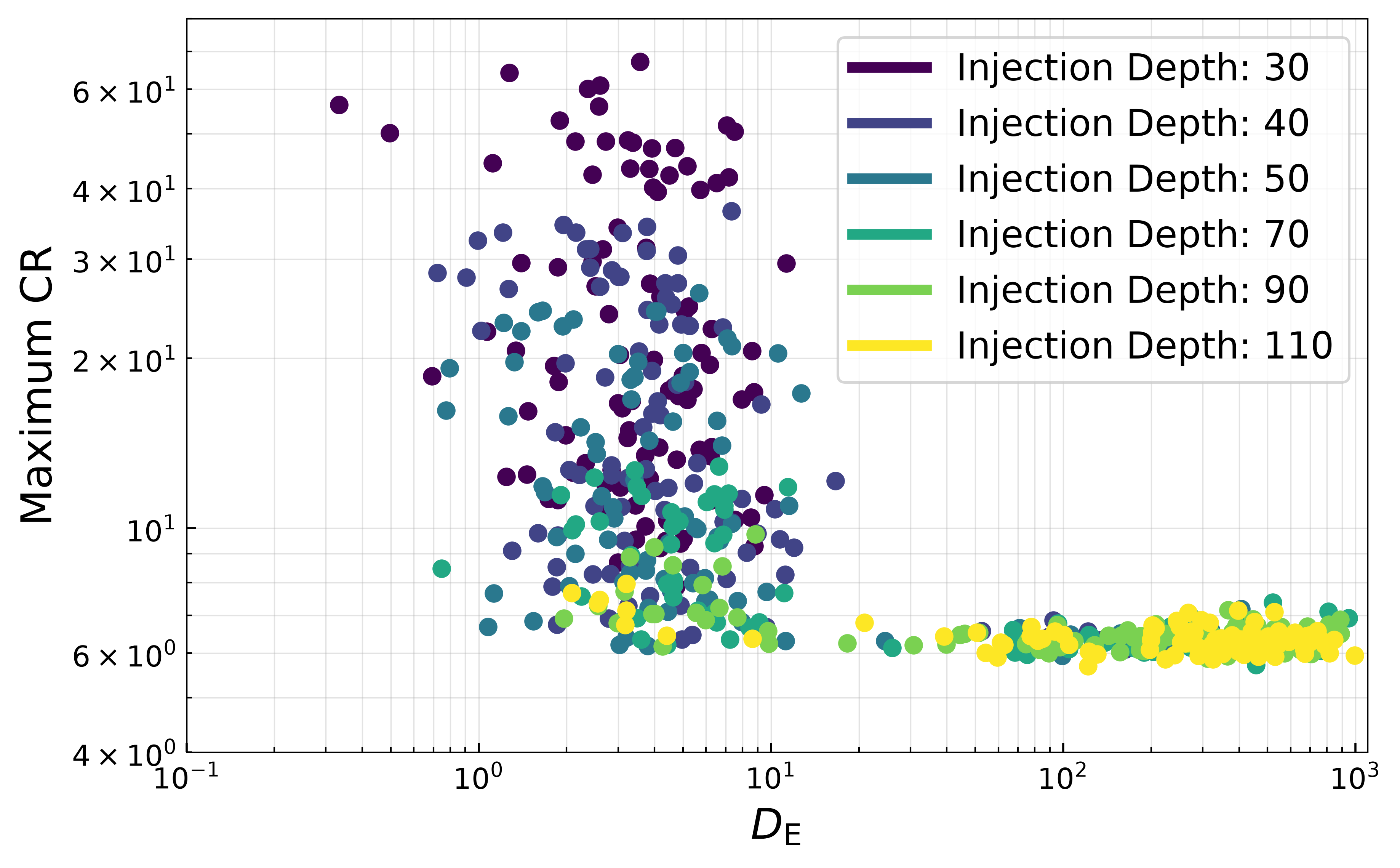}
    \includegraphics[width = 0.45\linewidth]{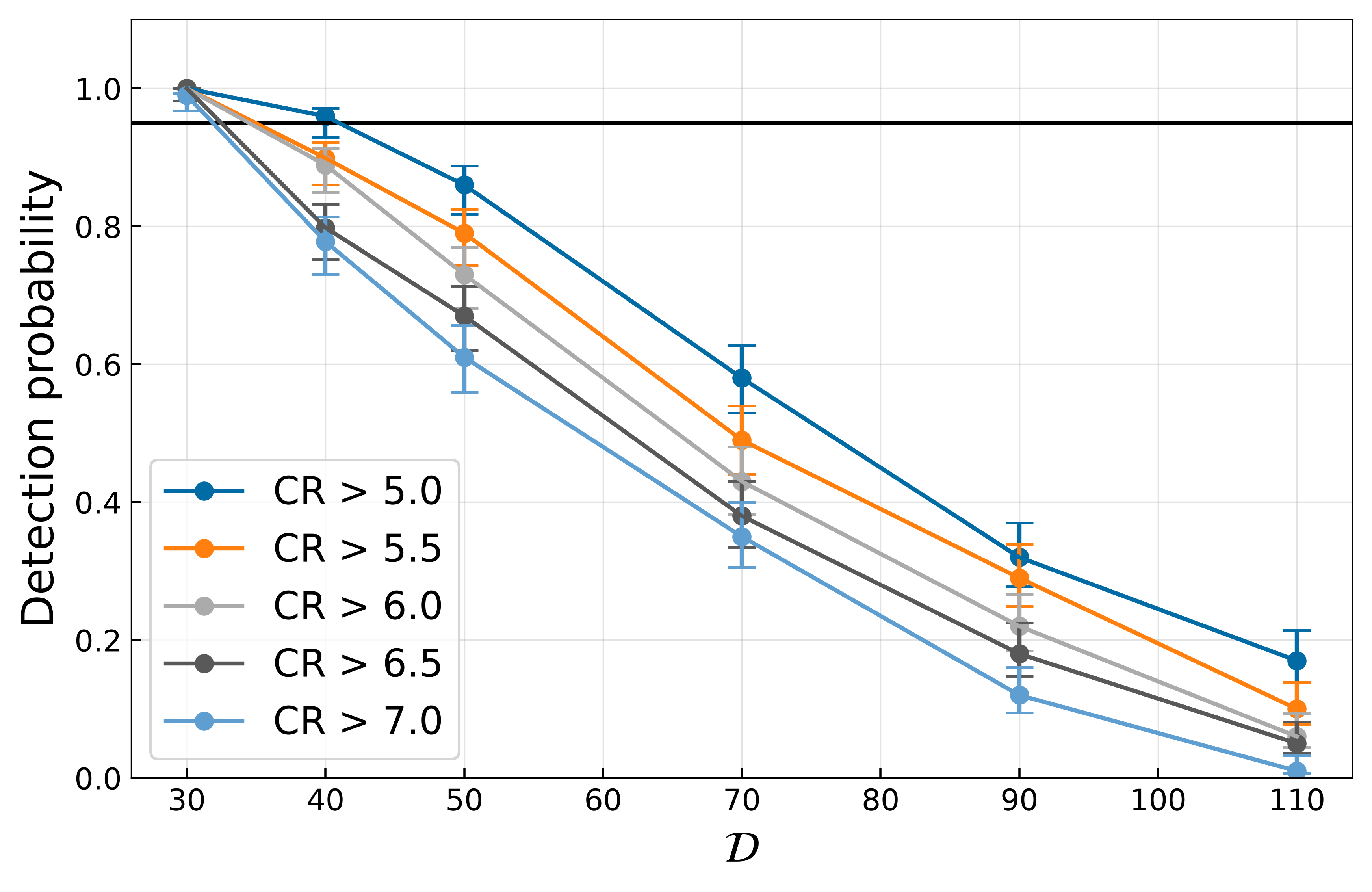}
    \includegraphics[width = 0.45\linewidth]{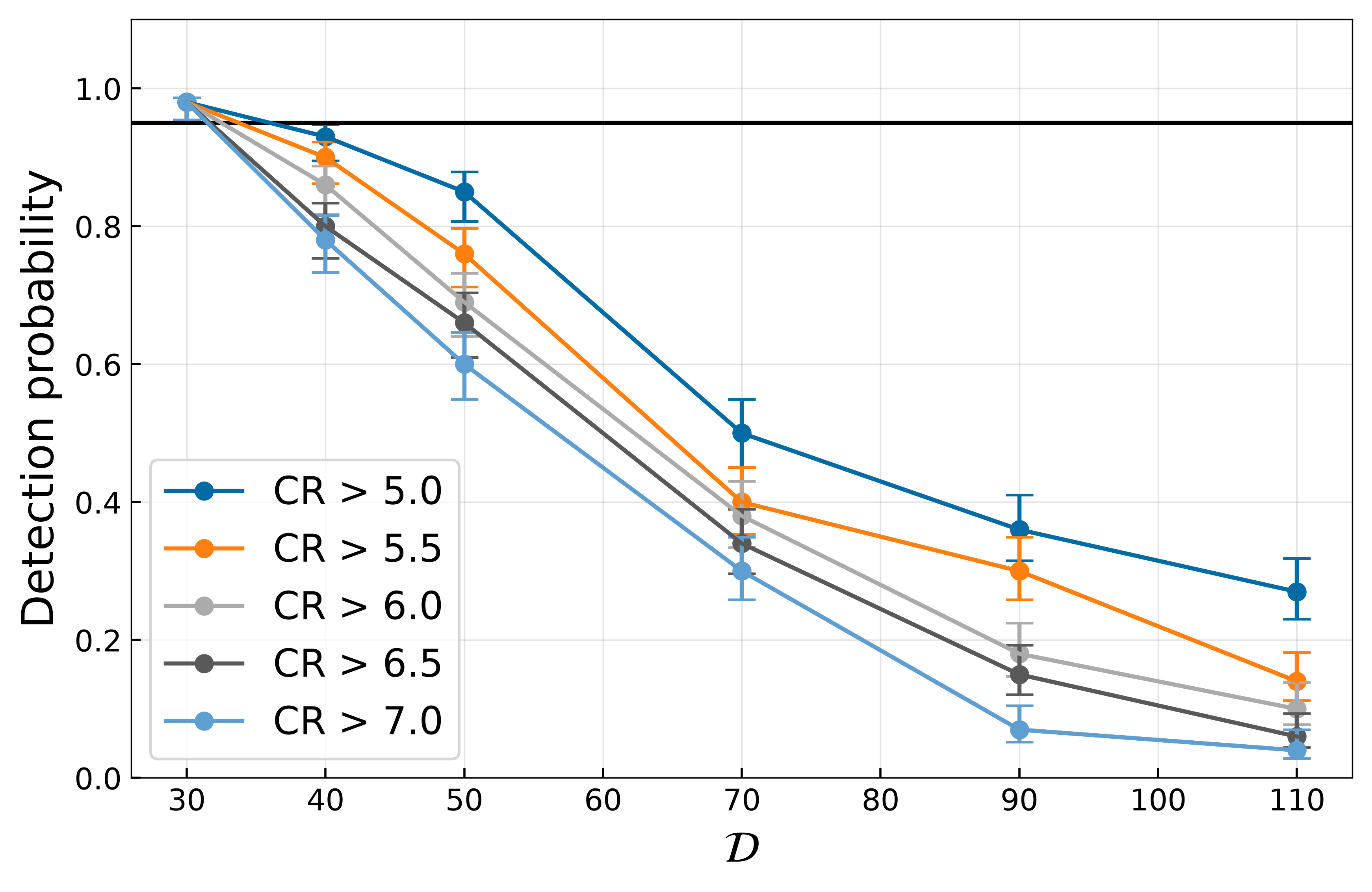}
    \caption{Same as Figure~\ref{fig:highf_injections_gaussian} but for software injections into real detector data.}
    \label{fig:highf_injections_real}
\end{figure*}

Figure~\ref{fig:highf_injections_real} shows injections into real detector noise for the same \gaia{} target as in Section~\ref{sec:real_noise_inj}.

